\documentclass[10pt]{sig-alternate-05-2015}
\usepackage{amsmath}
\usepackage{amsfonts}
\usepackage{amssymb}
\usepackage[T1]{fontenc}
\usepackage[nolist,nohyperlinks]{acronym}
\usepackage{enumitem}
\usepackage{graphicx}
\usepackage{cite}
\usepackage{url}
\usepackage{fancyvrb}
\usepackage{subfigure}
\usepackage{caption}
\usepackage{booktabs}
\setlist{nolistsep}
\usepackage{multirow}

\begin{document}

\CopyrightYear{2016} 
\setcopyright{acmlicensed}
\conferenceinfo{IMC 2016,}{November 14 - 16, 2016, Santa Monica, CA, USA}
\isbn{978-1-4503-4526-2/16/11}\acmPrice{\$15.00}
\doi{http://dx.doi.org/10.1145/2987443.2987445}

\title{Entropy/IP: Uncovering Structure in IPv6 Addresses}

\numberofauthors{3}
\author{
\alignauthor Pawe\l{} Foremski\\
	\affaddr{Akamai Technologies}\\
	\affaddr{IITiS PAN}\\	
    \email{pjf@iitis.pl}
\alignauthor David Plonka\\
	\affaddr{Akamai Technologies}\\
	\email{plonka@akamai.com}
\alignauthor Arthur Berger\\
	\affaddr{Akamai Technologies}\\
	\affaddr{MIT CSAIL}\\
	\email{arthur@akamai.com}
}
\maketitle

\begin{abstract}
In this paper, we introduce {\em Entropy/IP}: a system that discovers Internet address structure based on analyses of a subset of IPv6 addresses known to be active, {\em i.e.,} training data, gleaned by readily available passive and active means. The system is completely automated and employs a combination of information-theoretic and machine learning techniques to probabilistically model IPv6 addresses.
We present results showing that our system is effective in exposing structural characteristics of portions of the active IPv6 Internet address space, populated by clients, services, and routers.

In addition to visualizing the address structure for exploration,
the system uses its models to generate candidate addresses for scanning.
For each of 15 evaluated datasets, we train on 1K addresses and generate 1M candidates for scanning.
We achieve some success in 14 datasets,
finding up to 40\% of the generated addresses to be active.
In 11 of these datasets, we find active
{\em network identifiers} ({\em e.g.,} /64 prefixes or ``subnets'')
not seen in training.  Thus, we provide the first
evidence that it is practical to discover subnets and hosts by scanning
probabilistically selected areas of the IPv6 address space not known to contain
active hosts \textit{a priori}.

\end{abstract}

\begin{acronym}

\acro{P2P}{Peer-to-Peer}
\acro{CDN}{Content Delivery Network}
\acro{ISP}{Internet Service Provider}
\acro{DNS}{Domain Name System}
\acro{AI}{Artificial Intelligence}
\acro{PR}{Pattern Recognition}
\acro{ML}{Machine Learning}
\acro{DHT}{Distributed Hash Table}
\acro{BN}{Bayesian Network}
\end{acronym}

\section{Introduction} \label{sec:intro}
Understanding the structure of Internet addresses has become
increasingly complicated with the introduction, evolution, and operation of
Internet Protocol version 6 (IPv6).  Complications arise both from {\em (a)}
IPv6's address assignment features, {\em e.g.,} stateless address
auto-configuration (SLAAC),
in which clients choose their own addresses, and from {\em (b)} the freedom
allowed by IPv6's vast address space and enormous prefix allocations from
address registries,
{\em e.g.,} $2^{96}$ addresses (by default) to each Internet Service Provider (ISP).

By empirical observation today, we see many address complications in
the large-scale operation of IPv6.
As of August 2016, estimates suggest that 11\% of
World-Wide Web (WWW) users have IPv6 capability and use it to access
popular sites~\cite{GoogleIPv6}.
Yet, at this modest level, measurements
show {\em billions} of active IPv6 WWW client addresses being used monthly,
and tens to hundreds of millions of IPv6 router addresses.
Addresses often differ in the spatial and temporal characteristics
from one operator or network to the next~\cite{plonka2015temporal}.
Complications that we observed include (but are in no way limited to):
addresses with Modified EUI-64 interface identifiers that are curiously
{\em not} tagged as globally unique,
{\em stable} addresses containing pseudo-random numbers in their interface
identifiers,\footnote{\small The authors are aware of proposed
addressing schemes involving both randomized MAC addresses and stable
privacy addresses.  These privacy mechanisms increase complication.}
and even
addresses containing pseudo-random numbers in their {\em network identifiers}.
These are the sort of challenges we face with IPv6.

In this work, we study sets of IPv6 addresses and
the structural characteristics embedded within them.
We are primarily motivated by the challenges of widespread
native IPv6 operation (in parallel with IPv4), as it exists today.
First, WWW services often wish to deliver content differentially,
{\em e.g.,} based on a client's geographic location or its host reputation;
understanding IPv6 structure would help determine how these IPv4
characteristics might, likewise, apply to clients' IPv6 addresses.
Second, security analysts wish to be able to assess IPv6 networks'
vulnerability to host scanning; since full IPv6 address-space scans are
infeasible, new methods must take its structure into account.
Third, network operators, as well as engineers and researchers,
wish to track IPv6 deployment and growth to prioritize their work;
a systematic approach to deep understanding of IPv6 address structure aids
measurement interpretations that clearly differ from IPv4 to IPv6,
{\em e.g.,} so that we properly recognize the
significance of estimated subnet sizes and measured active host and
prefix counts.

The goal of our system, ``Entropy/IP,''
is to provide a means by which one might remotely glean and understand a network's addressing plan.
Ultimately, we would like to discover Classless Inter-Domain Routing (CIDR)
prefixes, Interior Gateway Protocol (IGP) subnets, network identifiers,
and interface identifiers (IID) of each address.
We employ three core techniques:

\begin{itemize} [wide]

\item {\bf Entropy Analysis:} A key aspect of our system is that it leverages
information-theoretic entropy~\cite{shannon2001mathematical}.
We employ it to measure the variability of values at each of
32 positions of {\em nybbles} ({\em i.e.,} hexadecimal characters) in
IPv6 addresses.
We then compare the entropy of adjacent nybbles to detect significant
differences, with the expectation that these represent boundaries between
semantically distinct parts of each address.
When there are no significant differences, we group adjacent
nybbles together to form larger {\em segments}.
For instance, subnet identifiers or stateless address auto-configuration
(SLAAC) attributes might be formed of adjacent nybbles that exhibit similar
entropy.

We chose to compute the entropy of nybbles at a certain position across a set of addresses
as an aid to understand the structure of these addresses.
Here, we are guided by our experience
that ``stateless'' analysis of individual address content is clearly
error-prone, and ``stateful'' analysis (spatial or temporal) has benefits
as Plonka and Berger~\cite{plonka2015temporal} show.
For instance, the reasonable, but stateless, rules to detect
pseudo-random IIDs implemented in the {\tt addr6} tool~\cite{ipv6toolkit},
misclassify
{\tt 2001:\-db8:\-221:\-ffff:\-ffff:\-ffff:\-ffc0:\-122a} as
having a randomized IID even when it is accompanied by one thousand
other similarly constructed addresses in the
{\tt 2001:\-db8:\-221:\-ffff:\-ffff:\-ffff:\-ff::/104} prefix.  Context is
important for accurate understanding of address content and entropy
is an ideal metric to recognize portions of addresses that
contain information that might be pertinent to address structure.
For instance, oft repeated bits---{\em e.g.,} those {\tt f}
characters---may contain hints about IID assignment practice, but
contain little {\em information} (in an information-theoretic sense)
and are, thus, unlikely to help in reverse-engineering netmasks and
subnet structure.

\item {\bf Clustering:} The second key aspect of our system is that we
employ unsupervised machine learning for clustering of the segments' values based on their distribution
and the frequencies of occurrence of those values.
We use the popular DBSCAN clustering algorithm~\cite{ester1996density}.

\item {\bf Statistical Modeling:} The third key aspect of our system
is the use of Bayesian Networks (BNs) to statistically model IPv6 addresses. In short, we automatically determine conditional probabilities amongst clusters of segments'
values in a hierarchical fashion, {\em i.e.,} directed left to right,
across the address segments.
This aspect was motivated by painstaking experiences in visual examination of large sets of 
IPv6 addresses.
It seems promising that a machine learning technique could unveil address
structure similar to how a researcher sometimes can.
Unsupervised learning seems especially appropriate since prior work found  
numerous instances of structure that does not follow RFC-defined
address assignment policies~\cite{plonka2015temporal}. Thus,
we do not train our system to recognize well-known features---such as
{\tt ff:fe} in Modified EUI-64, or ostensibly pseudo-random numbers
in privacy addresses---but rather rely on our system's
entropic underpinnings to {\em discover} these unique
characteristics for itself.
\end{itemize}

In stepwise fashion, Entropy/IP ingests a sample set of IP addresses,
computes entropies, discovers and mines segments, builds a BN model,
and prepares a graphical web page with the following elements for 
a network analyst to navigate and explore the exposed structure:
\begin{itemize} [wide]

\item a plot of entropy and aggregate count ratio,

\item a BN, showing address segments inter-dependencies,

\item a segment value browser with frequency heat map,

\item a target address generator.

\end{itemize}
Entropy/IP's user interface is shown in Figures~\ref{fig:example}
and~\ref{fig:example_struct}. A live demo, which allows for public operation, is available at \url{http://entropy-ip.com/}.

There are numerous applications of structural analysis of active IP
addresses.  These include {\em (a)} identifying homogeneous groups of
client addresses, {\em e.g.,} to assist in IP geolocation or in the
mapping of clients to content hosted on Content Distribution Networks (CDNs),
{\em (b)} supporting network situational awareness efforts, {\em e.g.,} in
cyber defense or in competitive analysis, and
{\em (c)} selecting candidate targets for active
measurements, {\em e.g.,} traceroutes campaigns, vulnerability assessments,
or reachability surveys.  With respect to survey, temporary addresses
complicate the estimation of users or service subscribers by counting
IPv6 addresses or prefixes at any one length ({\em e.g.,} 64 bits), so
understanding addressing in
network structure is critical in interpreting IPv6 address or prefix counts.
If we could interpret these counts, they could be used to inform,
{\em e.g.,} policy, standards, and capacity-planning decisions.
Yet another application 
of structure analysis is {\em (d)} remotely assessing networks'
addressing plan and address assignment policy. This is valuable
for host reputation and access control, {\em i.e.,} when mitigating
abuse originating from sources within that network.
Such external assessments are also valuable to the subject networks
{\em themselves,} {\em e.g.,} to assess potential security or privacy
risks~\cite{CLAB16}.
For instance, one network operator, whom we contacted to comment on our
results, asked {\em us} whether or not their customers' addresses could be
predicted, {\em i.e.,} whether or not their address assignments
appear to support user privacy as intended.

This work makes the following contributions:\\
{\em (1)} We present an automated system that discovers aspects of networks'
IPv6 address layout based on observations of a subset of that network's active
addresses.  Our system employs an entropy-based technique in combination with
standard machine learning and statistical modeling techniques to discover structural
characteristics in arbitrary IPv6 address sets.\\
{\em (2)} We improve upon prior works on address classification by employing
a measure of \emph{entropy} to identify address sets that have very high entropy values across multiple adjacent nybbles---which
likely reveal pseudo-random segments---and middle-to-high range values, as well as abrupt changes in entropy between segments---which
likely reveal addressing structure. This improves identification of privacy addresses.\\
{\em (3)} We present results and an evaluation of our system that
demonstrate how it enables analysts to interactively explore the
structural characteristics of arbitrary sets of addresses and, if desired,
generate candidate target addresses for active scanning more broadly
than existing methods described in the literature.

The remainder of this paper is organized as follows.  In
Section~\ref{sec:related}, we discuss related works.
In Section~\ref{sec:data}, we describe the data used in
our empirical study.  In Section~\ref{sec:methods}, we present our system, the methods by which it is implemented, and their component
techniques.  In Section~\ref{sec:eval}, we
present results of our evaluation.
In Section~\ref{sec:disc}, we discuss limitations of our
method and future work.  Subsequently, we conclude this paper in Section~\ref{sec:concl}.

\section{Related work} \label{sec:related}
To the best of our knowledge, the components in our method have not
previously been applied to the problem of uncovering IP address structure.
However, statistics, entropy, and machine learning
have been applied to network traffic analysis in numerous works
involving IP addresses and other traffic
features. For instance, Lee and Xiang~\cite{lee2001information} develop models for network anomaly
detection based on
entropy across features of traffic records, including host identifiers,
though they do not specifically mention IP addresses.
Feinstein {\em et al.}~\cite{feinstein2003statistical} develop a detector that relies on their observation that
attacks from distributed sources result in increased entropy of packet header
features; they focus on source IP address in their
evaluation.
Subsequently, both Lakhina {\em et al.}~\cite{Lakhina05} and Wagner {\em et al.}~\cite{wagner2005entropy} likewise
compute entropy across IP header features.
These works differ from ours in that they treat individual addresses as atomic,
{\em i.e.,}
semantically opaque, and largely use entropy as a measure of
feature distribution, {\em e.g.,} address set inflation, and
typically detect when it varies in time-series.

One work that deals specifically with entropy and IPv6 addresses is that
of Strayer {\em et al.}~\cite{strayer2004spie} Circa 2004, they note that the
IPv6 packet header, including the source and destination addresses, exhibits
less entropy (per byte) than that of IPv4.  We believe this is precisely
why our method is effective. In IPv6 addresses, information relevant
to structure or forwarding need not be ``compressed'' into only a 4-byte
identifier, as it is in IPv4 addresses.  Instead, information can (and often is) spread across an IPv6 address, {\em i.e.,} all of 32 nybbles.

Since then, the introduction and popularization of privacy
extensions~\cite{narten2007privacy} for IPv6 addresses changed the situation
by adding pseudo-random values, and thus high entropy, to IPv6 addresses.
While random values have high information content in an information-theoretic
sense, the information therein is not {\em pertinent} to network
structure. Thus, it is useful to identify these so-called ``privacy addresses''
and either disregard their random segments (IIDs) or treat them specially.
Works by Malone~\cite{DBLP:conf/pam/Malone08},
Gont and Chown~\cite{gont2016network} (as implemented in the {\tt addr6}
tool~\cite{ipv6toolkit}), and Plonka and Berger~\cite{plonka2015temporal} each
attempt to identify pseudo-random numbers in IPv6 address segments.
We do so as well in this work, differing in that we leverage entropy.

Kohler {\em et al.}~\cite{DBLP:conf/imc/KohlerLPS02} employ
a hierarchical approach within IPv4 addresses and count aggregates
(prefixes)
at each possible length, 0 through 32.
Plonka and Berger~\cite{plonka2015temporal} build on that approach
and introduce Multi-Resolution Aggregate (MRA) Count Ratios for IPv6 addresses
and use them to discover structure in addresses.
They were also inspired by prior works~\cite{DBLP:conf/qofis/ChoKK01,DBLP:conf/lisa/EstanM05} that summarize the IP address space
into an abbreviated structure, albeit entirely
synthetic.
In contrast, in this work we take a looser hierarchical approach: we separately consider each nybble position and we group them into segments of varying lengths. Thus, Entropy/IP provides a complimentary viewpoint into IPv6 addresses, independent from MRA analysis. However, for the interest of readers who are familiar with MRA plots, we show 4-bit Aggregate Count Ratios (ACR) in some of our figures (normalized to a range of $0-1$). At a high level, ACR reveals how much a segment of the address is relevant to grouping addresses into areas of the address space. The higher the ACR value, the more pertinent to prefix discrimination a given segment is. To understand the contribution of this paper, one can ignore the ACR metric, though.

Krishnamurthy and Wang's work~\cite{krishnamurthy2000network}, circa 2000, is similar to ours in its
premise, {\em i.e.,} automated grouping of homogeneous active addresses,
where only the BGP prefixes are known in advance. It differs significantly
in its methods and focuses only on IPv4.

Nascent work, which at first glance is most similar to ours, is that of 
Ullrich {\em et al.}~\cite{ullrich2015reconnaissance}, who develop a pattern-discovery-based scanning approach to IPv6 reconnaissance.
They algorithmically detect recurring bit
patterns ({\em i.e.,} structure) in the IID portion of training subsets of
known router, server, and client addresses, and then generate candidate
targets according to those patterns.
They report improved performance versus
the surveillance approach outlined by Gont and
Chown~\cite{gont2016network} that relies on a mixture of
patterns known or observed {\em a priori}, as implemented in the
{\tt scan6}~\cite{ipv6toolkit} tool.  Our work differs most significantly
from these prior works in
that we discover patterns across {\em whole} IPv6 addresses, including the
network identifier portion, whereas theirs focus only on the bottom 64 bits ({\em i.e.,} ostensibly the IID). Thus, they assume a 
surveyor or adversary knows which /64 prefixes to target.
Since our method can also be used to generate target /64 prefixes (see Section~\ref{sec:predict_prefix}),
it could be used in concert with theirs.

There are a number of informative related works regarding applications
of our system, {\em e.g.,} active scanning and probing of the IP address space.
They are pertinent in that {\em (a)} they offer the performance necessary
for scanning at large scale~\cite{zmap13} or {\em (b)}
they find that address discovery by passive monitoring significantly
improves target selection, and therefore efficiency and/or coverage, in
subsequent active scanning campaigns~\cite{DainottiCCR2014,CLAB16}.
Most recently, Gasser {\em et al.}~\cite{gasser2016scanning} focused
specifically on the challenge of generating hit lists for scanning the IPv6
address space based on IPs gleaned from a number of common sources, some
of which we use as well. They also contribute IPv6 support in {\tt zmap}~\cite{zmap13},
suggesting IPv6 scanning is feasible, but do not contribute a
strategy to algorithmically generate candidate targets.
The Shodan search engine resorted to infiltrating
the NTP server pool to discover active IPv6 host addresses that
it subsequently probed~\cite{Shodan2016}, presumably since sequential or
random full scanning of the IPv6 space is infeasible.
Our work differs from these in that we probabilistically generate hit lists of
targets not yet seen.

\section{Datasets} \label{sec:data}
For our study, we used 3.5 billion IPv6 addresses total, all of which were collected in Q1 2016 from several data sources. We assembled both small sets for various real-world networks and large sets, aggregated by type. Table \ref{tab:datasets} summarizes the smaller datasets of 3 types: Servers, Routers, and Clients (end-users). In each category, we arranged IPv6 addresses (IPs) into individual sets for each of 5 major operators.

\begin{table}[htb]
\centerline{\includegraphics[scale=0.7,clip=true,trim=10mm 148mm 106mm 12mm]{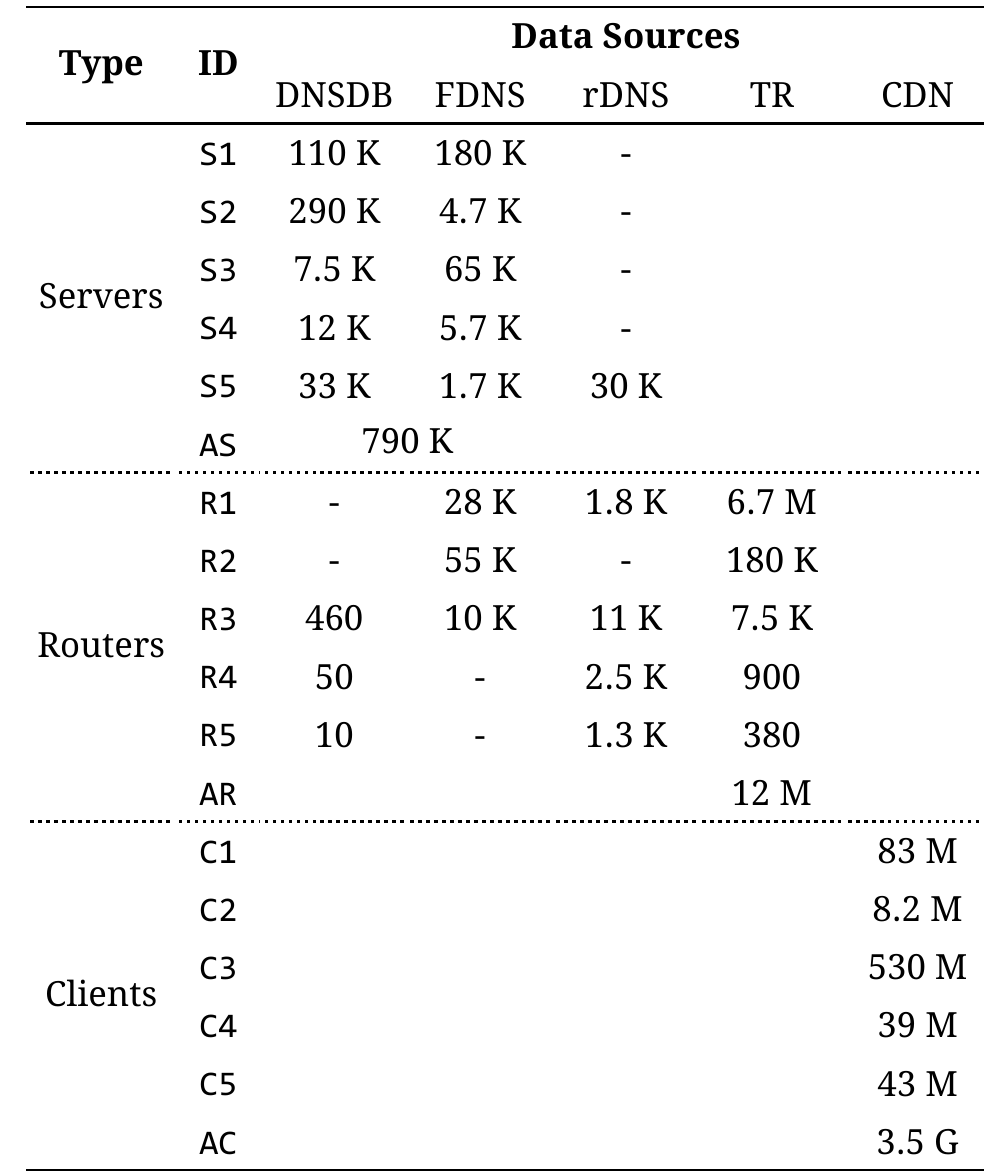}}
\caption{Number of unique IPv6 addresses in ``Small'' datasets of IPv6 addresses (\texttt{S*}, \texttt{R*}, and \texttt{C*}), and in ``Aggregate'' datasets (\texttt{AS}, \texttt{AR}, and \texttt{AC}).} \label{tab:datasets} \vspace{-0.5em}
\end{table}

For Servers (\texttt{S1}-\texttt{S5}), \texttt{S1} represents a web hosting company, \texttt{S2} and \texttt{S3} represent two different CDNs, and \texttt{S4} represents a certain cloud provider. The operator of \texttt{S5} is commonly known for offering all of these services, and for providing many public web services, \emph{e.g.}, a social network.
Among Routers (\texttt{R1}-\texttt{R5}), all datasets represent router interfaces of major global Internet carriers. For Clients (\texttt{C1}-\texttt{C5}), all datasets represent leading ISP networks that deliver wired and mobile Internet access for domestic and enterprise customers.

To collect Server addresses, we employed the Domain Name System (DNS), as it is most common to ren\-dez\-vous with services by domain name. For Routers, we used DNS and a large-scale traceroute dataset (column ``TR'' in Table \ref{tab:datasets}), comprised of router interface addresses on paths between servers of a major CDN, and from these servers to some clients. For Clients, we used addresses of the clients involved in web requests to the CDN, during 17-23 March 2016. In order to evaluate with generally available data, we also collected client addresses via the BitTorrent network.

The first data source in Table \ref{tab:datasets} (column ``DNSDB'') presents the number of addresses found via DNSDB: a large DNS database offered by Farsight Security \cite{dnsdb} that passively collects DNS data worldwide. It offers a broad view of queries and responses, allows for fetching forward records by host or network address (an ``inverse'' query), and resolves wild-card queries (\emph{e.g.}, \texttt{*.ip6.\-isp.\-net/\-PTR}). For Servers, we queried DNSDB for prefixes used by operators for their IPv6 services, inferred from WHOIS and BGP data. For Routers, besides prefixes, we used wild-carded forward and reverse domain queries. We restricted the router IPs gleaned from DNS to those that appeared in our traceroutes.

For the second data source (column ``FDNS''), we applied analogous techniques on the Forward DNS dataset by Rapid7 Labs \cite{fdns}. This dataset is periodically recreated by actively querying DNS for domains found in various sources (including TLD zone files, Internet-wide scans, web pages, etc). For the last DNS data source (column ``rDNS'') we applied the technique described in RFC 7707~\cite[pp.~23]{gont2016network} by Gont and Chown that leverages DNS reverse mappings for obtaining IPv6 addresses of a particular network.

We also collected \emph{aggregate} datasets for each category: \texttt{AS} for Servers, \texttt{AR} for Routers, and \texttt{AC} for Clients. As data sources, we used DNS for \texttt{AS} (790K IPs in 4.3K /32 prefixes), large-scale traceroute measurements for \texttt{AR} (12M IPs in 5.5K /32 prefixes), and 7-day CDN traffic for \texttt{AC} (3.5 billion IPs in 6.0K /32 prefixes). The aggregates cover the individual sets presented in Table~\ref{tab:datasets}. In order to avoid some networks from being over-represented, in Section \ref{sec:aggregates}, we used stratified sampling by randomly selecting 1K IPs from the /32 prefixes.

Inspired by work of Defeche and Vyncke \cite{defeche2012measuring}, we also collected an aggregate of client addresses from the public BitTorrent \ac{P2P} network: dataset \texttt{AT}. To collect these, we built a custom BitTorrent client that crawled various trackers and \ac{DHT} peers during 5-8 March 2016. We collected 220K peer addresses in 1.8K /32 prefixes by running our software from Singapore, US East Coast, and Europe.

We employ address anonymization when presenting results. We changed the first 32 bits in IPv6 addresses to the documentation prefix (\texttt{2001:db8::/32}), incrementing the first nybble when necessary. To anonymize IPv4 addresses embedded  within IPv6 addresses, we changed the first byte to the \texttt{127.0.0.0/8} prefix.

\vspace{0.6em}

\section{Methodology} \label{sec:methods}
\begin{figure}[h!]
\centerline{\includegraphics[scale=0.5,clip=true,trim=0mm 30mm 26mm 3mm]{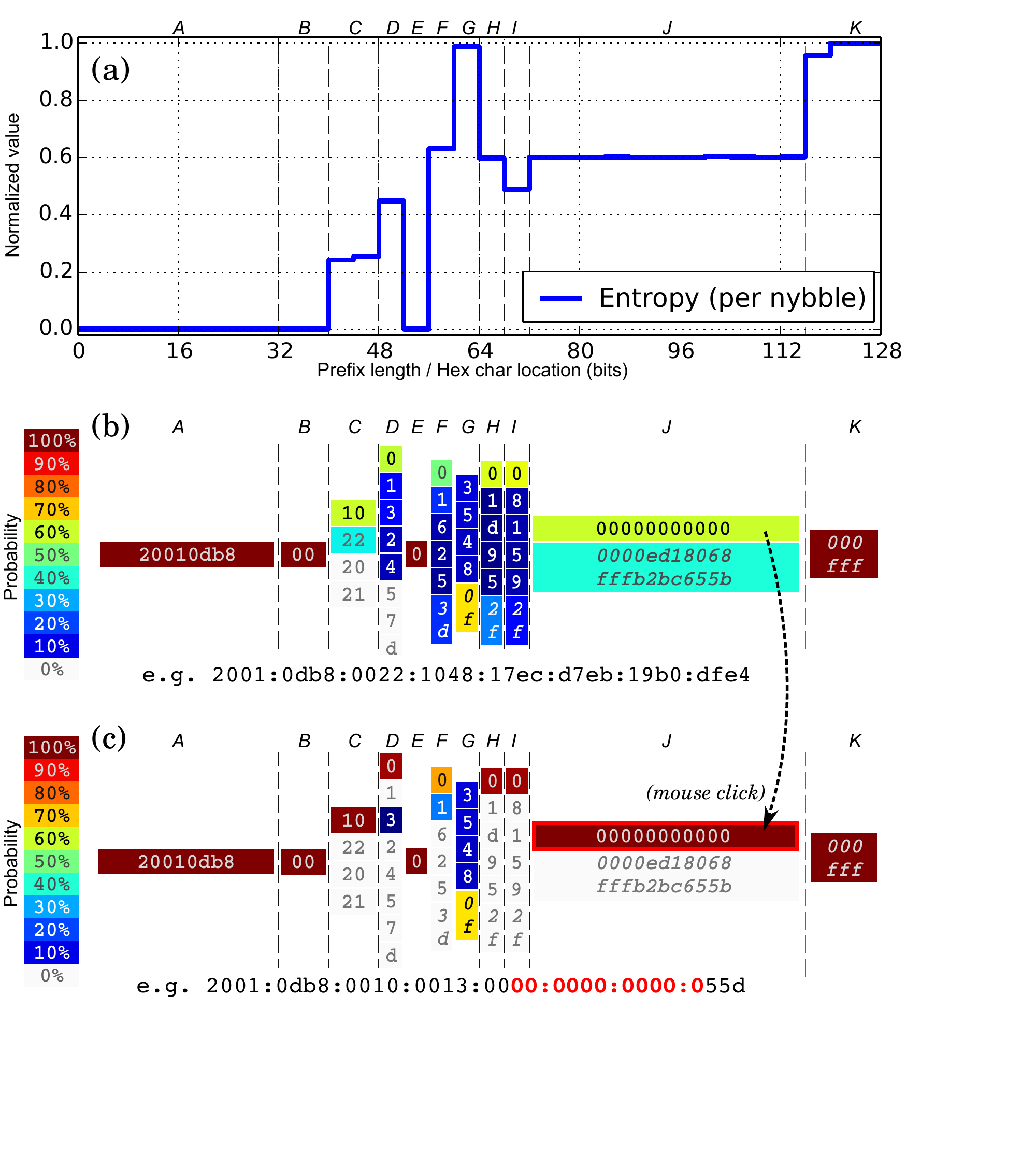}}
\caption{Entropy/IP's user interface displaying an analysis of a Japanese telco prefix with 24K active client IPs. Entropy by nybble plotted in (a). In (b), we select the {\tt 00000$\dots$} value (60\%) for segment {\em J} by mouse click, resulting in updated probabilities in (c) ({\em e.g.,} 100\%).}
\label{fig:example}

\vspace{1.5em}

\centerline{\includegraphics[scale=0.23,clip=true,trim=0mm 1mm 0mm 0mm]{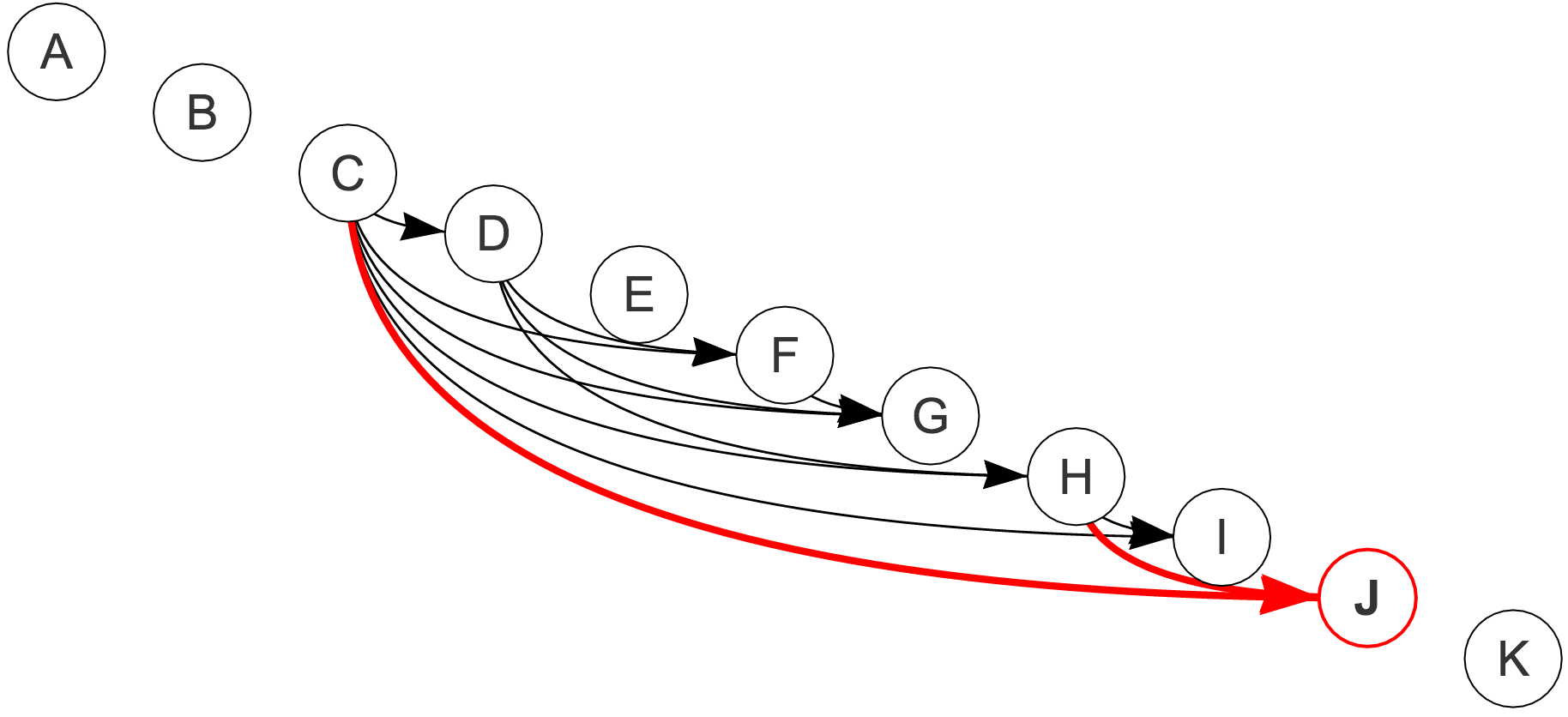}}
\caption{Dependencies between segments. Red color indicates direct probabilistic influence on segment J.}
\label{fig:example_struct}

\vspace{0.5em}

\begin{center}
\begin{tabular}{c|cccc}
\toprule
\multirow{2}{*}{H} & \multicolumn{4}{|c}{C} \\
 & \texttt{10} & \texttt{22} & \texttt{20} & \texttt{21} \\
\midrule
\texttt{0} & 100\% & 0.48\% & 7.7\% & 14\% \\
\texttt{1} & 17\% & 0.17\% & 6.3\% & 13\% \\
\texttt{d} & 25\% & 0.17\% & 7.7\% & 5.9\% \\
\texttt{9} & 17\% & 0.17\% & 9.1\% & 9.1\% \\
\texttt{5} & 20\% & 0.17\% & 7.1\% & 7.1\% \\
\emph{\texttt{2-f}} & 2.3\% & 0.017\% & 0.79\% & 0.77\% \\
\bottomrule
\end{tabular}
\captionof{table}{Probability for segment {\em J} in Fig. \ref{fig:example_struct} equal {\tt 00000$\dots$}, conditional on values in segments {\em H} and {\em C}.}
\end{center}
\label{tab:example_probs}
\end{figure}

In this section, we introduce our system by its visual interface, and then we detail our underlying methodology. In Fig.~\ref{fig:example}, we present the analysis results for a set of 24K WWW client addresses in a Japanese telco's prefix, collected from a CDN during a week's time.

The main components of Entropy/IP's visual interface are as follows. First, Fig.~\ref{fig:example}(a) plots entropy per address nybble, across the dataset. (We detail this in Section~\ref{sec:entropy}.) Here, the trained eye can see that the addresses are covered by one /40 prefix. In short, the address segments---delineated by dashed vertical lines and labeled with capital letters {\em A} through {\em K} at the top---are comprised of nybbles having similar entropy. Apart of that, we always make bits 1-32 the segment \emph{A}. (We detail this in Section~\ref{sec:segments}.)

Second, Fig.~\ref{fig:example}(b,c) are examples of Entropy/IP's conditional probability browser. Here, we show the distributions of values inside segments by a colored heat map. (We detail this in Section~\ref{sec:mining}.)
For example, segment {\em A} always has the value {\tt 20010db8}, which is reflected in 100\% probability. In this example, the length of segment {\em C} is two nybbles, in which four distinct values were observed: the most popular being {\tt 10} at 60\% in Fig.~\ref{fig:example}(b). Ranges are shown as two values (low to high) within one colored box, {\em e.g.,} segment {\em J} having an interval of {\tt 0000ed18068} to {\tt fffb2bc655b} at 40\%.

In the transition from Fig.~\ref{fig:example}(b) to Fig.~\ref{fig:example}(c), the analyst is curious how the probabilities
would change if one conditioned on the segment {\em J} having the value {\tt 00000$\dots$}.
Clicking on this value yields Fig.~\ref{fig:example}(c), showing for instance that now {\em C} has the value {\tt 10} at 100\%, and likewise for value {\tt 0} in segments {\em H} and {\em I}.

Fig.~\ref{fig:example_struct} shows the structure of an associated Bayesian Network (BN), with nodes representing the segments and edges indicating a statistical dependency. (We detail this in Section~\ref{sec:bn}.)
Here, the red edges show that the segment {\em J} is directly dependent on segments {\em C} and {\em H}, which is analyzed in Table~\ref{tab:example_probs}. The segments can influence each other in the opposite direction and through other segments. Thus, selecting a particular value for \emph{J} influences \emph{F} through \emph{C}, which is the reason for different distribution for the segment \emph{F} in Fig.~\ref{fig:example}(c) vs. Fig.~\ref{fig:example}(b).

A live, functional demo of Entropy/IP interface is publicly available at \url{http://entropy-ip.com/}.

\vspace{0.6em}

\subsection{Entropy Analysis} \label{sec:entropy}

\emph{Entropy} is a measure of unpredictability in information content \cite{shannon2001mathematical}. Usually it is defined as $H(X)$ for a discrete random variable $X$ with possible values $\lbrace x_1, \cdots, x_k \rbrace$ and a probability mass function $P(X)$:
\begin{align} \label{eq:entropy}
H(X) = -\sum^{k}_{i=1} P(x_i) \log{P(x_i)}.
\end{align}

\noindent In general, the higher the entropy, the more equally probable values of $X$ are: if $H(X)=0$, then $X$ takes only one value; if $H(X)$ is maximum, then $P(X)$ is uniform. For the remainder of the paper, we normalize entropy by dividing it by $\log{k}$ (maximum value).

\begin{figure}[h!]
\centerline{\includegraphics[scale=0.65,clip=true,trim=0mm 0mm 0mm 0mm]{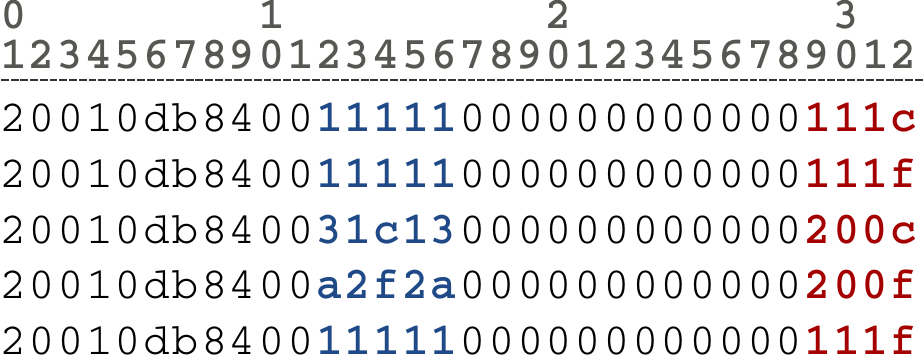}}
\caption{Sample IPv6 addresses in fixed-width format, sans colon characters.\label{fig:sampleHexIp}}
\end{figure}

In order to use entropy for analyzing the structure of IPv6 addresses, let $D$ be a set of addresses expressed as 32 hexadecimal digits without colons, \emph{e.g.}, as in Fig.~\ref{fig:sampleHexIp}.\vspace{-0.75em}
 \vspace{-0.75em}

\noindent Let us consider the values in the $i$-th hex character position of the addresses in $D$ as instances of a random variable $X_i$, $i=1,\dots,32$. 
That is, let us focus on values of a specific nybble across all addresses in $D$. In Fig.~\ref{fig:sampleHexIp}, the last character takes ``\texttt{c}'' twice and ``\texttt{f}'' thrice. Thus, $X_{32}$ has empirical probability mass function $\hat{P}(X_{32}) = \lbrace p_c=\frac{2}{5}, p_f=\frac{3}{5} \rbrace$. Since there are 16 possible hex characters, the maximum entropy is ${\log{16}}$, and thus
the normalized empirical entropy is:
\begin{align}
\hat{H}(X_{32}) = \frac{-(p_c\log{p_c} + p_f\log{p_f})}{\log{16}} \approx 0.24.
\end{align}

By repeating the above for all $i$, we get 32 values that reveal statistical properties of individual hex characters across the set: the smaller the value, the greater chances the character stays constant. Let us also introduce a notion of \emph{total entropy} $\hat{H}_S$, \vspace{-0.5em}
\begin{align}
\hat{H}_S(D) = \sum^{32}_{i=1} \hat{H}(X_i),
\end{align} \vspace{-0.5em}

\noindent which quantifies variability of addresses in $D$, {\em i.e.,} how hard it is to guess actual addresses by chance\footnote{\small Note: ``total entropy'' is not the entropy if one considered the whole address as a single element, and computed the probability mass function of those elements---that entropy, normalized, would be very low.}.

\subsection{Address Segmentation}
\label{sec:segments}
Entropy exposes the parts of IPv6 addresses that are variable versus those that remain relatively constant. Let us use it to group adjacent hex characters into contiguous blocks of bits with similar entropy. We will call such blocks \emph{segments} and label them with capital letters. For instance, in Fig.~\ref{fig:sampleHexIp}, the hex characters 1-11 and 17-28 are constant (entropy $=0$), whereas the values in hex characters 12-16 and 29-32 are changing (entropy~$\neq 0$). Hence, they form 4 segments: A (1-11), B (12-16), C (17-28), and D (29-32). By segmenting IPv6 addresses, we distinguish contiguous groups of bits that differ in joint variability.

We propose a simple threshold-based segmentation algorithm. Consider the entropy of successive nybbles, left to right. Start a new segment at nybble $i$ whenever $\hat{H}(X_i)$ compared with $\hat{H}(X_{i-1})$ passes through any of the thresholds $T = \lbrace 0.025,\- 0.1,\- 0.3,\- 0.5,\- 0.9 \rbrace$.
We also employ a hysteresis of $T_h = 0.05$, \emph{i.e.}, we require
\begin{align}
|\hat{H}(X_i) - \hat{H}(X_{i-1})| > T_h
\end{align}
to start the new segment. For example, if $\hat{H}(X_{i-1})=0.49$, then in order to start the next segment $\hat{H}(X_i)$ has to be either ${<}0.3$ or ${>}0.54$, with $0.3$ being the lower threshold for $\hat{H}(X_{i-1})$ in $T$ (without hysteresis) and $0.54$ being $\hat{H}(X_{i-1})+T_h$ (with hysteresis.) We found this set of parameters $T$ and $T_h$ during development by evaluation on real-world networks. The parameters can be tuned to match specific networks, yet we identified the proposed values to be universal and produce least number of segments with similarly distributed nybbles.

In addition to the thresholds, we always make the bits 1-32 a separate segment. This is motivated by the common practice of Regional Internet Registries (RIRs), who use a /32 prefix as the smallest block assigned to local Internet operators~\cite{arin32}. Similarly, we always put a boundary after the 64\textsuperscript{th} bit, as it commonly separates the network identifier from the interface identifier \cite{hinden2006ip}.

\subsection{Segment Mining}
\label{sec:mining}
Further in our analysis, we want to understand why some segments appear non-random (\emph{i.e.}, have entropy~${<}1$). We hope to uncover common elements of IPv6 addresses, which possibly have a semantic meaning \cite{jiang2013analysis}.

Let us delve into a specific segment $k$. First, reduce the input dataset $D$ down to $D_k$: for each address, drop all nybbles outside of segment $k$. Next, search $D_k$ for the set $V_k$ of popular values and ranges that cover considerable parts of $D_k$. For example, in Fig.~\ref{fig:sampleHexIp} the segment of nybbles 12-16 has $D_k = \lbrace$\texttt{11111}, \texttt{11111}, \texttt{31c13}, \texttt{a2f2a}, \texttt{11111}$\rbrace$ and $V_k = \lbrace \texttt{11111} \rbrace$.

We propose a heuristic approach for building $V_k$ that focuses on three aspects of data: {\em (a)} frequencies of values, {\em (b)} the values themselves, and {\em (c)} both characteristics considered together. We address them separately in the steps described below. In each step, we nominate at most the top 10 elements to $V_k$ and remove them from $D_k$. If there is ${\leq}0.1\%$ of values left, we finish.

\begin{figure}[htb]
\centerline{\includegraphics[scale=0.45,clip=true,trim=0mm 0mm 0mm 0mm]{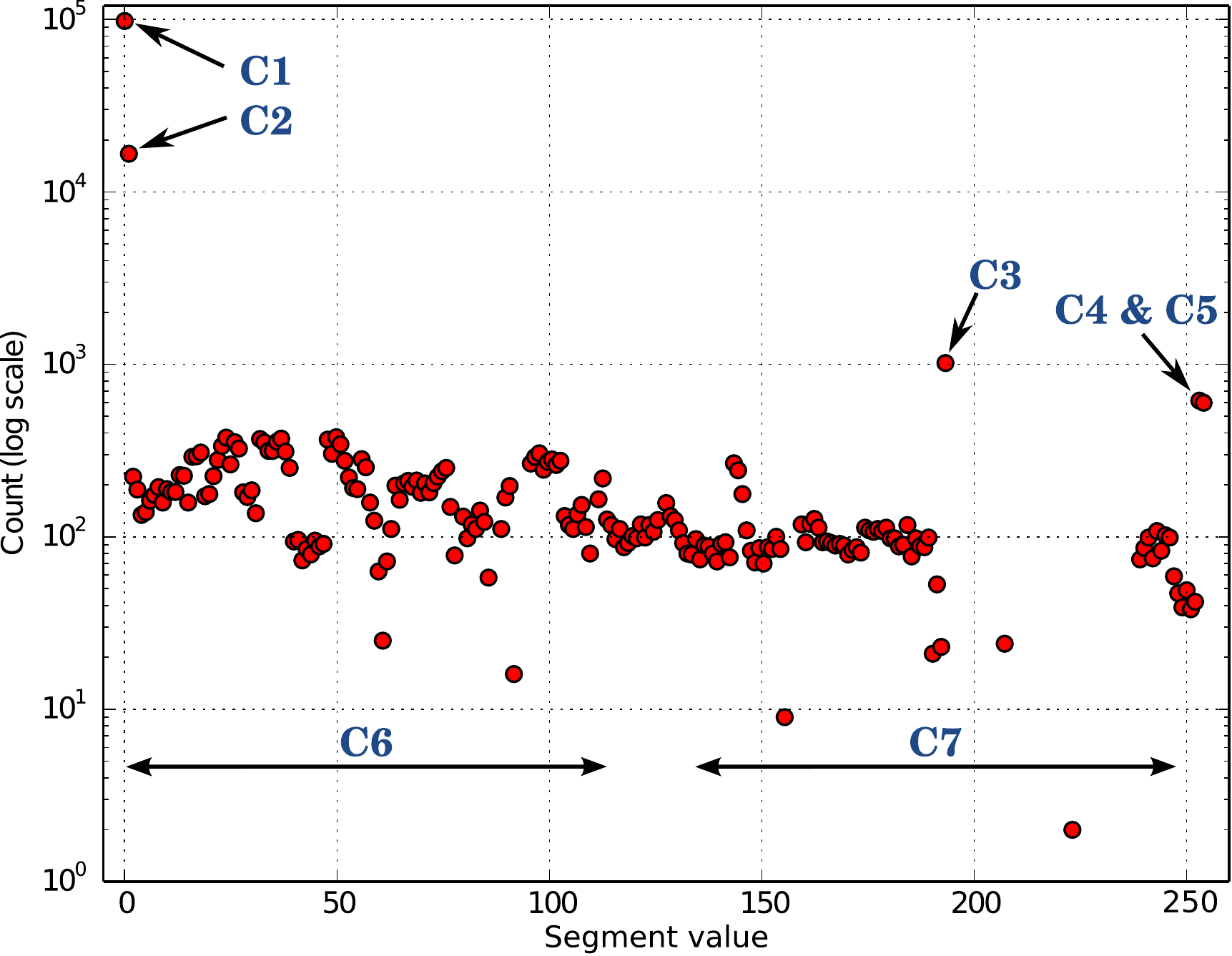}}
\caption{Histogram for values in segment C of dataset \texttt{S1}, as a scatter plot. Annotations show codes of common values and ranges.}\label{fig:histo} \vspace{-1em}
\end{figure}

For {\em (a),} we use a well-known outlier detection method to find unusually prevalent values in $D_k$.  Assuming normal distribution of \emph{frequencies} of values, we select the values more common than $Q_3+1.5{\cdot}IQR$, where $Q_3$ is the third quartile and $IQR$ is the inter-quartile range. For example, see the values labeled C1 through C5 in Fig. \ref{fig:histo}. 
In this example, the segment has length of two nybbles (thus 256 possible values), which is the X-axis in Fig. \ref{fig:histo}, and
the Y-axis is the number of times a given value appears in set $D_k$.
Next, for {\em (b),} we run on $D_k$ the popular DBSCAN data clustering algorithm \cite{ester1996density}, parametrized to find highly dense ranges of values. In this step, we use the minimum and maximum values of the discovered clusters as ranges added to $V_k$. Then, for {\em (c),} we run the DBSCAN algorithm again, but on a histogram of $D_k$, that is, on a vector of values vs. their counts. We tune the algorithm to find ranges of values that are both uniformly distributed and relatively continuous (\emph{e.g.}, C6 in Fig. \ref{fig:histo}). Finally, if anything is left, we either close $V_k$ with a range of $(\min{D_k}, \max{D_k})$, or if $|D_k| \leq 10$ we take the whole $D_k$.

\begin{table}[tb]
\centerline{\includegraphics[scale=0.58,clip=true,trim=13mm 47mm 63mm 12mm]{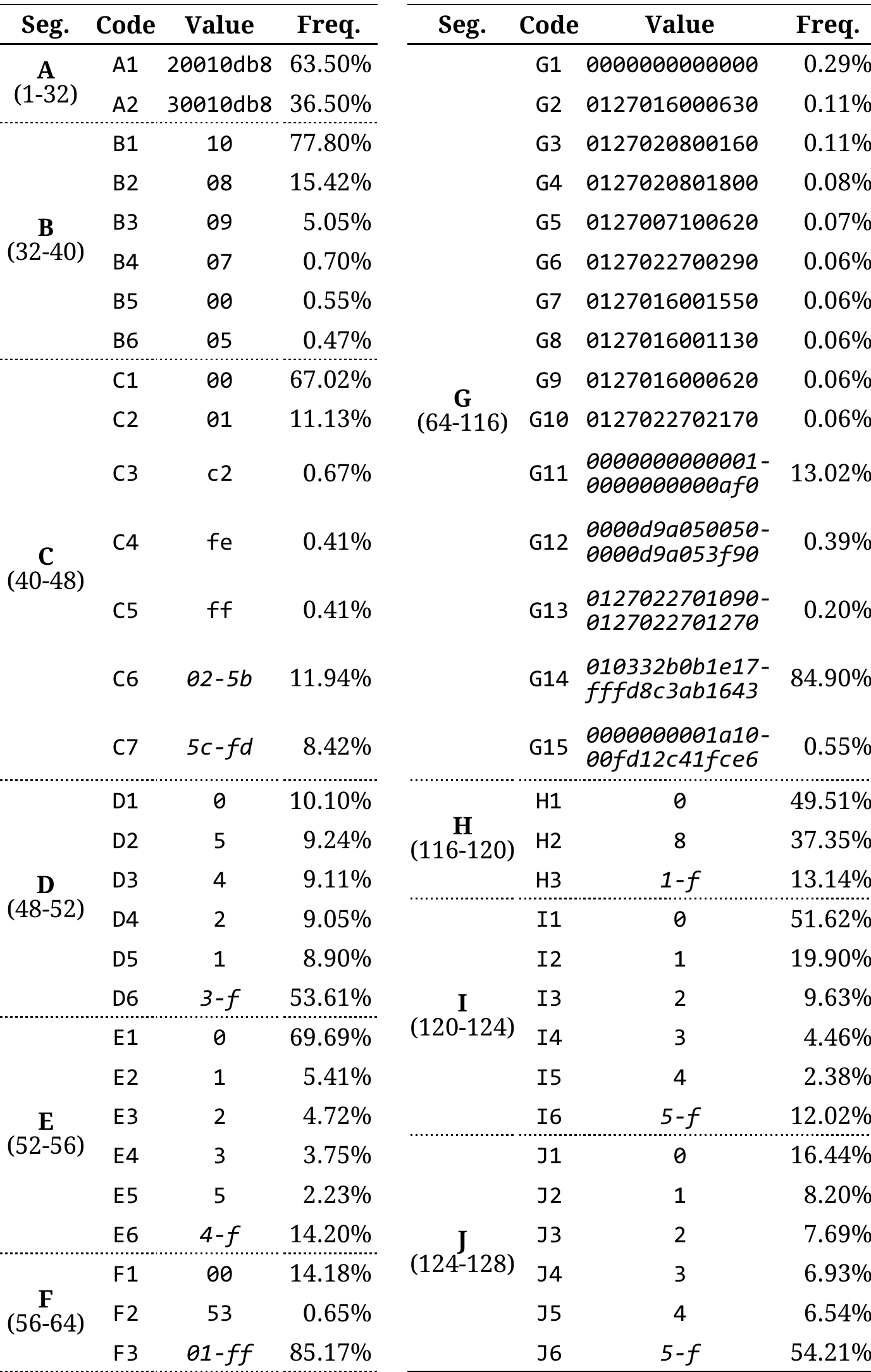}}
\caption{Segment mining results for dataset \texttt{S1}.}
\label{tab:analysis} \vspace{-1.7em}
\end{table}

We preserve the order of elements added to $V_k$ and we keep their empirical frequencies. In Table \ref{tab:analysis}, we present the result of our algorithm for the \texttt{S1} dataset. The column ``Code'' gives labels for segment values, which can be used to encode (or ``compress'') addresses, {\em e.g.,}
\begin{center}
\begin{BVerbatim}
2001:0db8:08c2:2500:0000:d9a0:5345:0012
\end{BVerbatim}
\end{center}
can be rewritten using Table \ref{tab:analysis} as a vector:
\begin{center} \vspace{-1.8em}
\begin{equation*}
( \text{A1, B2, C3, D4, E5, F1, G12, H1, I2, J3} ).
\end{equation*}
\end{center} 
\noindent Note that we lose details of the original address when we encode a segment with a code that represents a range (\emph{e.g.}, G12); this is acceptable for our purposes. In further discussion we represent IPs as instances of random vectors, where each dimension corresponds to segment $k$ and takes categorical values that reference $V_k$.

\subsection{Modeling IPv6 Addresses}
\label{sec:bn}
Having the addresses represented as random vectors enables us to easily apply well-known statistical models. Consequently, we can reveal inter-dependencies between the segments, and uncover structures hidden within IPv6 addresses. For this, we use Bayesian Networks (BNs).

BN is a statistical model that represents jointly distributed random variables in the form of a directed acyclic graph \cite{koller2009probabilistic}. Each vertex represents a single variable $X$ and holds its probability distribution, conditioned on the values of the other variables. An edge from vertex $Y$ to $X$ indicates that $X$ is statistically dependent on $Y$. BN can be used to model complex phenomena involving many variables. It splits complex distributions into smaller, interconnected pieces, which are easier to comprehend and manage.

Let us find a BN that represents a dataset of IPv6 addresses rewritten as random vectors. We need to learn the BN structure from data ({\em i.e.,} discover statistical dependencies), and we need to fit its parameters ({\em i.e.,} estimate conditional probability distributions). For this purpose, we use the ``BNFinder'' software, which implements the relevant state-of-the-art methods \cite{dojer2006learning,wilczynski2009bnfinder}. Since learning BNs from data is generally NP-hard, we constrain the network so that given segment $k$ can only depend on previous segments ${<}k$, \emph{e.g.}, $B$ can directly depend on $A$, but not on $C$. However, note that $C$ can still influence $B$ through evidential reasoning; that is, probabilistic influence can flow ``backwards.'' We refer the reader to \cite{koller2009probabilistic} for more details on BNs.

Once the BN model is found, we may use it for multiple purposes. For example, we may query the BN with various segment values set \emph{a priori} and discover how this affects the probability distributions in the other segments. We may also use the BN to generate candidate addresses that match the model (optionally constrained to certain segment values), which can be used for targeted scanning of IPv6 networks.

\subsection{Discussion}
We presented a heuristic system that, given a set of IPv6 addresses, discovers address segments with their popular values and a probabilistic structure. Obviously, we did not fully explore the design space possible for such a system; thus, our viewpoint is neither definitive nor the only possible. In order to encourage further development of similar tools, below we comment on our design choices and on possible adaptations.

We chose to focus on 4-bit chunks of IPv6 addresses, \emph{i.e.}, nybbles. It is possible to adapt Entropy/IP to other bit widths by modifying the entropy analysis step. For instance, one could use the bit widths of 1 or 16, as used by Plonka and Berger in \cite{plonka2015temporal} for MRA plots. However, we found the 4-bit approach simple and sufficient. If more granularity is needed, then the segment mining step can discover individual values. Conversely, if less granularity is needed, then the address segmentation step can coalesce adjacent nybbles. Besides, the 4-bit granularity matches the textual representations of IPv6 addresses, in which a single hex character is the smallest possible address chunk, thus matches the human analyst's canonical perspective.

\begin{figure}[t!]
\centerline{\includegraphics[scale=0.46,clip=true,trim=0mm 0mm 0mm 0mm]{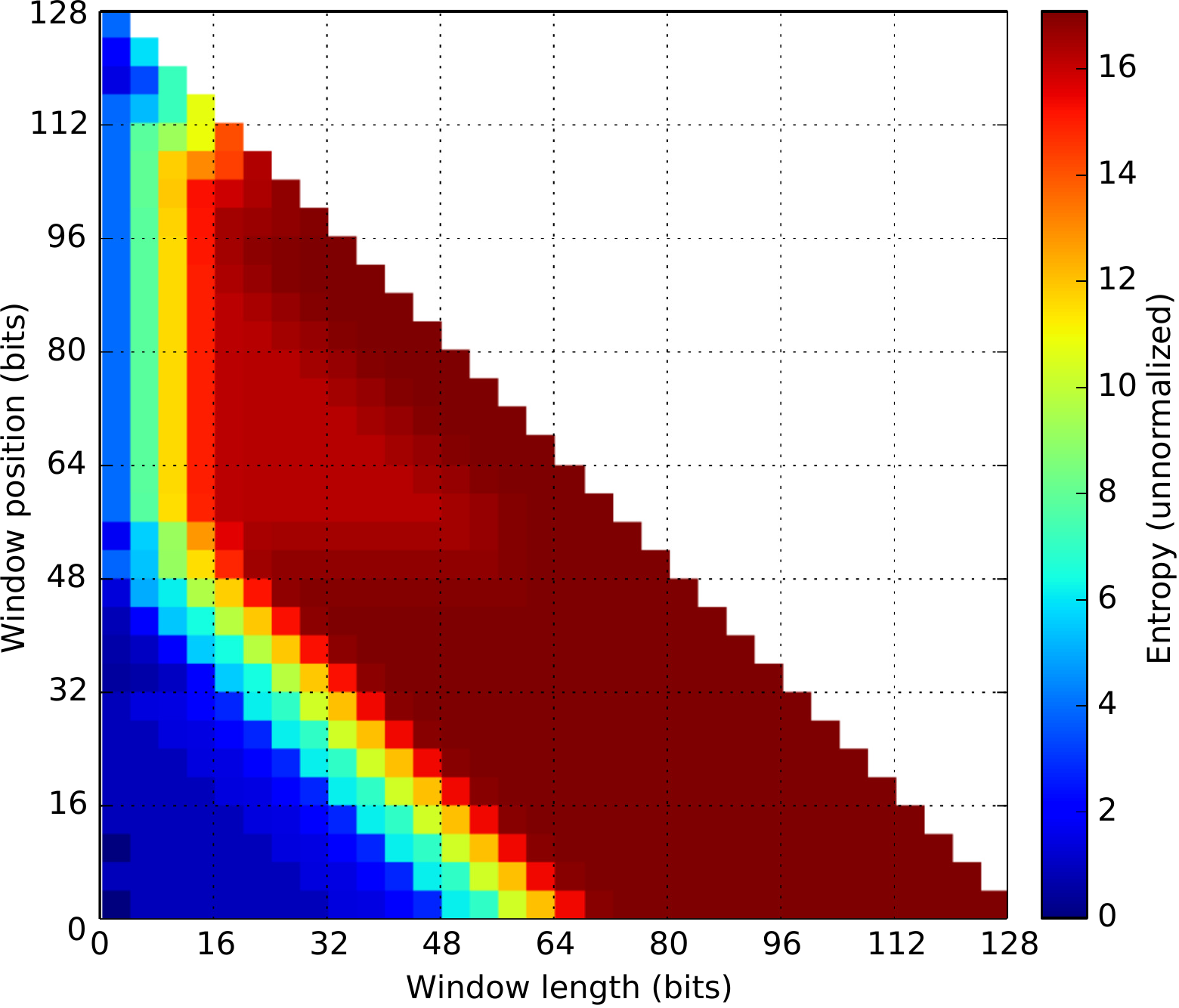}}
\vspace{-0.5em} \caption{An illustration for the preliminary idea of windowing analysis of entropy (dataset \texttt{S1}).}
\label{fig:windowing} \vspace{-1em}
\end{figure}

Let us briefly mention a preliminary idea of \emph{windowing analysis}, devised during early development of Entropy/IP. For a set of IPv6 addresses, we evaluated the entropy (unnormalized) for every possible address segment, determined by windows of varying length and position. For example, in Fig.~\ref{fig:windowing} we visualize such an analysis for dataset \texttt{S1}: every $(X,Y)$ point on the plane shows the entropy calculated across the dataset for bits $Y$ through $Y+X$. Here, note that one could use a different variability measure than the entropy, \emph{e.g.}, number of distinct values, inter-quartile range, frequency of the most popular value, or a weighted mean thereof. We believe this may be especially useful in conjunction with the windowing analysis for visual discovery of patterns.

We also tried segmenting using the difference between entropy of adjacent nybbles. However, taking into account the number of segments and distributions of their nybbles, the simple thresholds algorithm performed better. Note that segmentation without \emph{a priori} knowledge is problematic. Judging just by the apparent features of an addressing space may lead to uncovering the intents of network administrators, but may impose an artificial model on the data as well. However, we believe the segment mining and BN modeling steps together can compensate for minor glitches in segmentation.

Finally, we considered other tools (than BNs) for modeling the IPv6 addresses, \emph{e.g.}, Probability Trees (PTs)~\cite{drake1967fundamentals} and Markov Models (MMs)~\cite{rabiner1986introduction}. PTs, although conceptually simple, require information on virtually every possible combination of the segment values, and hence need abundant training data. The other alternative, MMs, assume that a given segment depends only on the previous segment. Thus, MMs cannot directly handle dependency between non-adjacent segments. Overall, we found BNs to produce models that are powerful, easy to explore, and concise.

\section{Evaluation} \label{sec:eval}
In this section, we demonstrate the efficacy of our methods on the datasets introduced in Section \ref{sec:data}.

\subsection{Big Picture: Aggregates} \label{sec:aggregates}

\begin{figure}[t!]
\centerline{\includegraphics[scale=0.46,clip=true,trim=10mm 4mm 15mm 13mm]{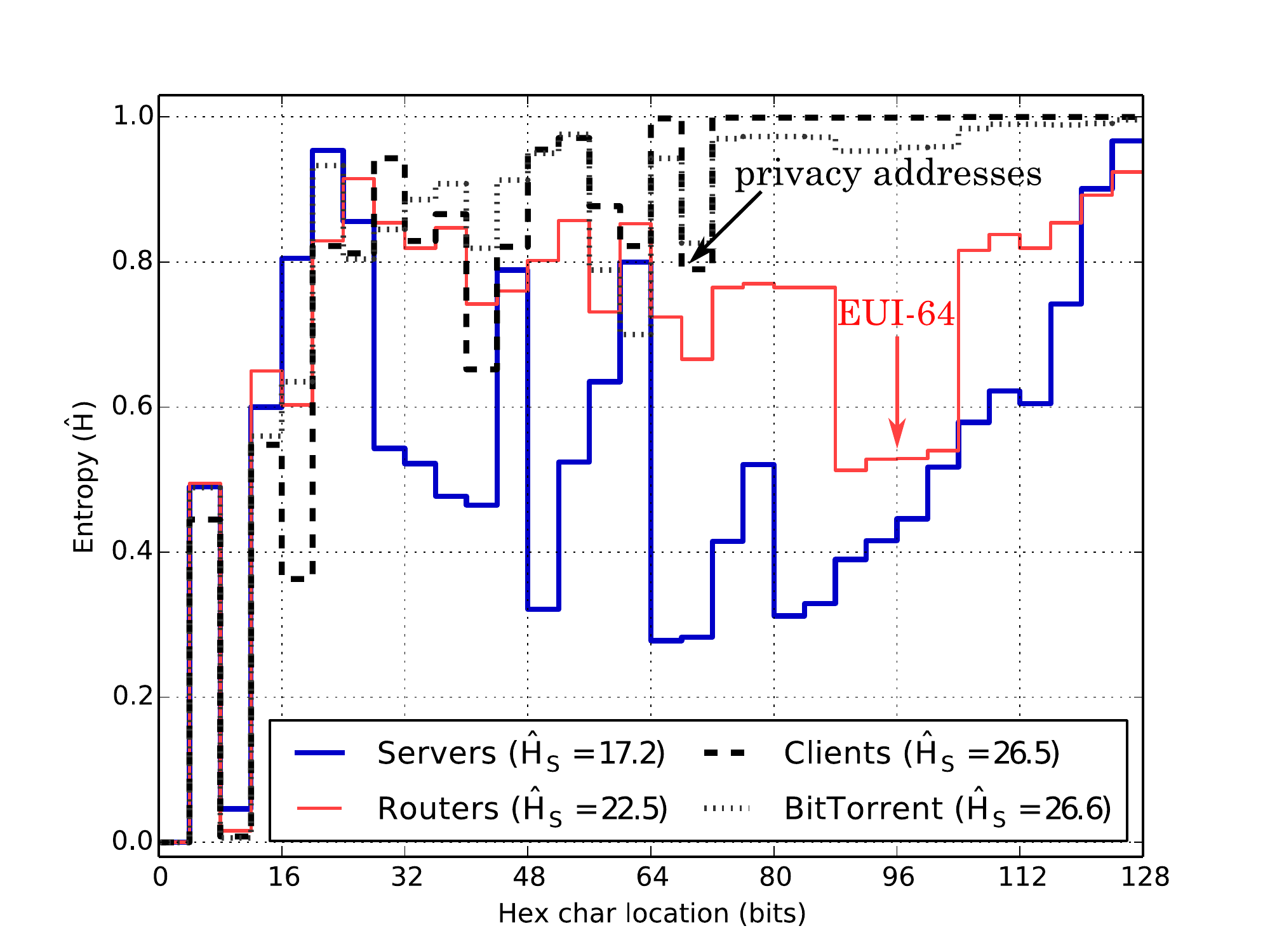}}
\caption{Entropy of aggregate datasets (\texttt{A*}).}
\label{fig:aggregates}
\end{figure}

\begin{figure*}[tb]
    \centering
    \subfigure[entropy vs. 4-bit ACR]{\includegraphics[scale=0.51,clip=true,trim=10mm 68mm 16mm 10mm]{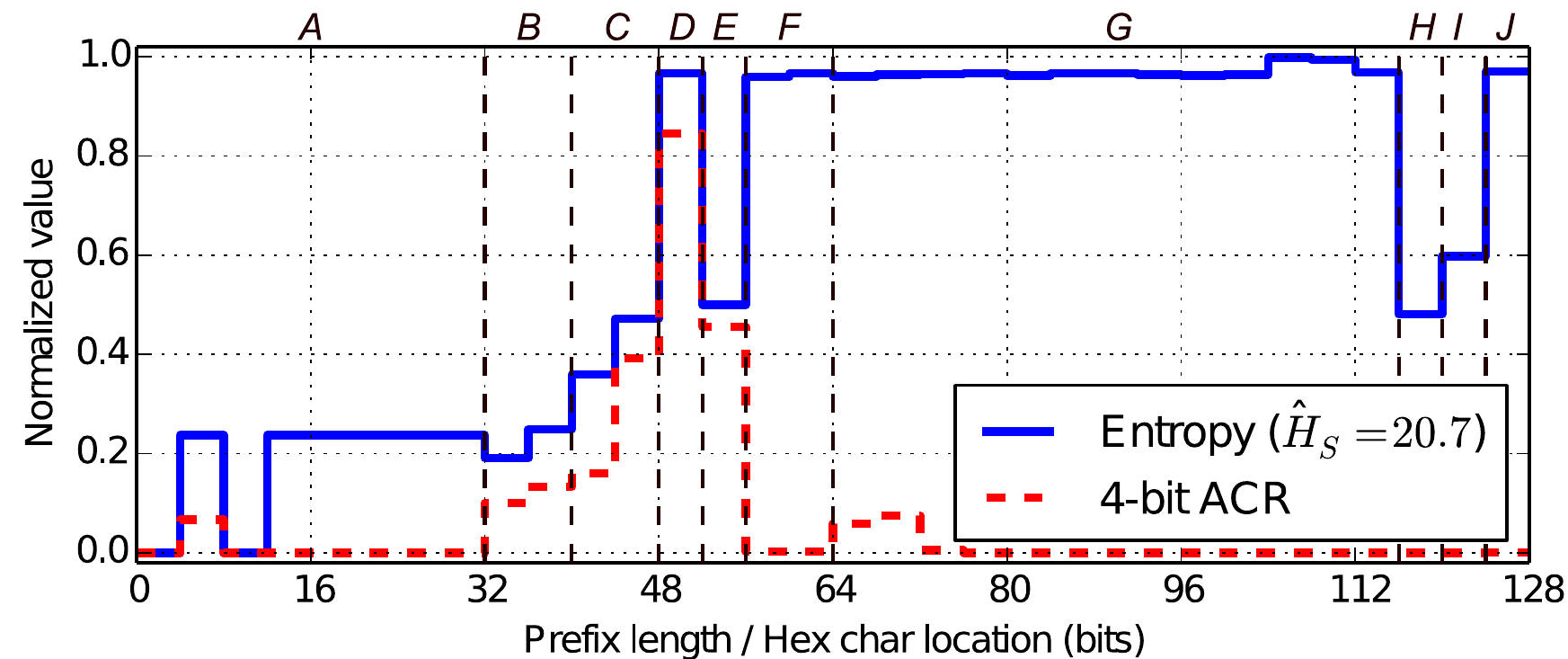} \label{fig:s1a}}
	\subfigure[BN probabilities conditional on B equal \texttt{08} or \texttt{09} (skipped probabilities ${<}0.1\%$ in G for brevity)]{\includegraphics[scale=0.88,clip=true,trim=59mm 236mm 57mm 17mm]{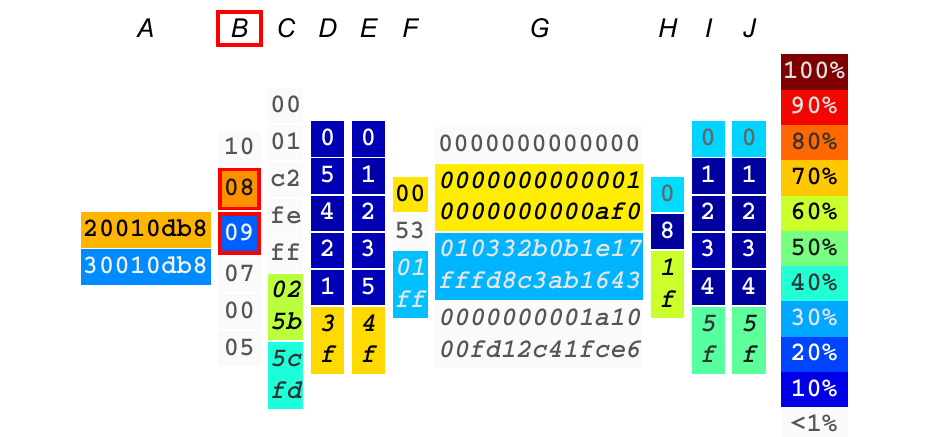} \label{fig:s1b}} \vspace{-0.5em}
\caption{Results for server dataset \texttt{S1}.} \vspace{-0.5em} \label{fig:s1} 
\end{figure*}

In Fig. \ref{fig:aggregates}, we present entropy characteristics for IPv6 addresses in our aggregate datasets, for three types of hosts. We see that the characteristics differ considerably past the first 32 bits; this is reflected in the $\hat{H}_S$ metric. Among the datasets, the client addresses were the most random, especially in the bottom 64 bits, ostensibly the IPv6 interface identifier. We see $\hat{H} \approx 1$, with the exception of $\hat{H} \approx 0.8$ for bits 68-72. This largely matches the RFC specifications for the ``u'' bit in SLAAC addresses~\cite{narten2007privacy}. However, if all of the clients used temporary addresses, we would observe an entropy of $0.75$ for bits 68-72. Hence, some of these addresses were not standard SLAAC privacy addresses.

We see a similar yet smaller dip for router addresses in bits 68-72, and a deeper drop to $\hat{H}\approx 0.5$ in bits 88-104. This suggests impact of IPv6 addresses based on link-layer identifiers (e.g., Ethernet MAC address), which have the word \texttt{0xfffe} inserted in bits 88-104 \cite{hinden2006ip}. Again, if all of the router addresses followed this standard, we would observe $\hat{H}=0$. Hence, a major portion of router addresses did not have MAC-based Modified EUI-64 IIDs.

On the other hand, in Fig.~\ref{fig:aggregates}, we see no evidence of Modified EUI-64 or privacy SLAAC addresses for servers. Instead, we highlight an interesting phenomenon: the entropy is oscillating across the address, revealing popular locations for discriminators of various entities in the
addressing hierarchy ({\em e.g.,} subnetworks). In general, the addresses in dataset \texttt{AS} are the least random, compared with the other sets. This demonstrates that servers are more likely to have predictable addresses. Note the steady increase in entropy from bit 80 to 128; this reflects the tendency to use the lower order bits when assigning static addresses to servers.

Finally, in Fig.~\ref{fig:aggregates}, we do not find significant differences between client addresses collected from CDN (dataset \texttt{AC}) vs. collected from the BitTorrent network (dataset \texttt{AT}), except for bits 88-104, which suggests Modified EUI-64 SLAAC addressing is more common for BitTorrent clients than general web clients. Thus, we believe it is possible to glean structure in client IPv6 networks using such freely-available data sources.

\subsection{Servers}

In Fig.~\ref{fig:s1}, we present analysis results for dataset \texttt{S1}---the addresses of a major web hosting provider---in which we discovered 10 segments. The BN model, visualized in Fig.~\ref{fig:s1b}, shows probability distributions across these segments, conditioned on {\em B} equal to either {\em B2} or {\em B3} (real values \texttt{08} or \texttt{09}, respectively), which represents approx. $20\%$ of addresses in \texttt{S1}.

The network has two /32 prefixes, which differ in popularity: 64\% vs. 36\%. Their actual values are anonymized in Fig.~\ref{fig:s1b}, but in Fig.~\ref{fig:s1a} we see in segment {\em A} that the two actual prefix values differ in six hex characters; their entropy is non-zero. ACR is non-zero only for bits 4-8, which means each /8 prefix holds just one /32 prefix. By exploring the BN model, we found the addressing scheme largely the same for both prefixes, but {\em B1} is 10\% more likely for {\em A2} than for {\em A1}.

We further found that segment {\em B} selects a variant of addressing used on the lower bits: probability distributions differ for {\em B1} vs. {\em B2}/{\em B3}, vs. {\em B4}/{\em B6}, vs. {\em B5}. In other words, the network has 4 variants of addressing deployed across its /40 prefixes. For example, in Fig.~\ref{fig:s1a} we see high entropy in segments {\em F} and {\em G} ({\em 
i.e.,} high variability in bits 56-116). However, when we constrained the segment {\em B} to {\em B2}/{\em B3}, in Fig.~\ref{fig:s1b}, we find a major drop in the variability of bits 56-116: the majority of addresses in this variant are essentially non-random. For the \emph{B4/B6} variant, we found 67\% of IPv6 addresses encode literal IPv4 addresses in segments \emph{G-J}. We verified these IPv4 addresses belong to the same operator and respond to ping requests with similar round-trip times as the IPv6 addresses in which they were embedded. (This strongly suggests these IPv6/IPv4 address pairs are aliases on dual-stacked hosts.)

Next, in Fig.~\ref{fig:s1a}, we see that segments {\em C} through {\em E} have high values for both entropy and ACR. This ostensibly means that bits 40-56---apart from being variable---are utilized for discriminating prefixes. In contrast, for segment {\em F} (bits 56-64), we see high entropy with ACR near zero. In this area, the address is variable, but appears to carry little useful information to discriminate addresses from one another; typically, each /56 prefix, here, covers just a single active {\em random} /64 prefix.
\begin{figure*}[htb]
    \centering
	\subfigure[servers]{\includegraphics[scale=0.51,clip=true,trim=16mm 4mm 72mm 10mm]{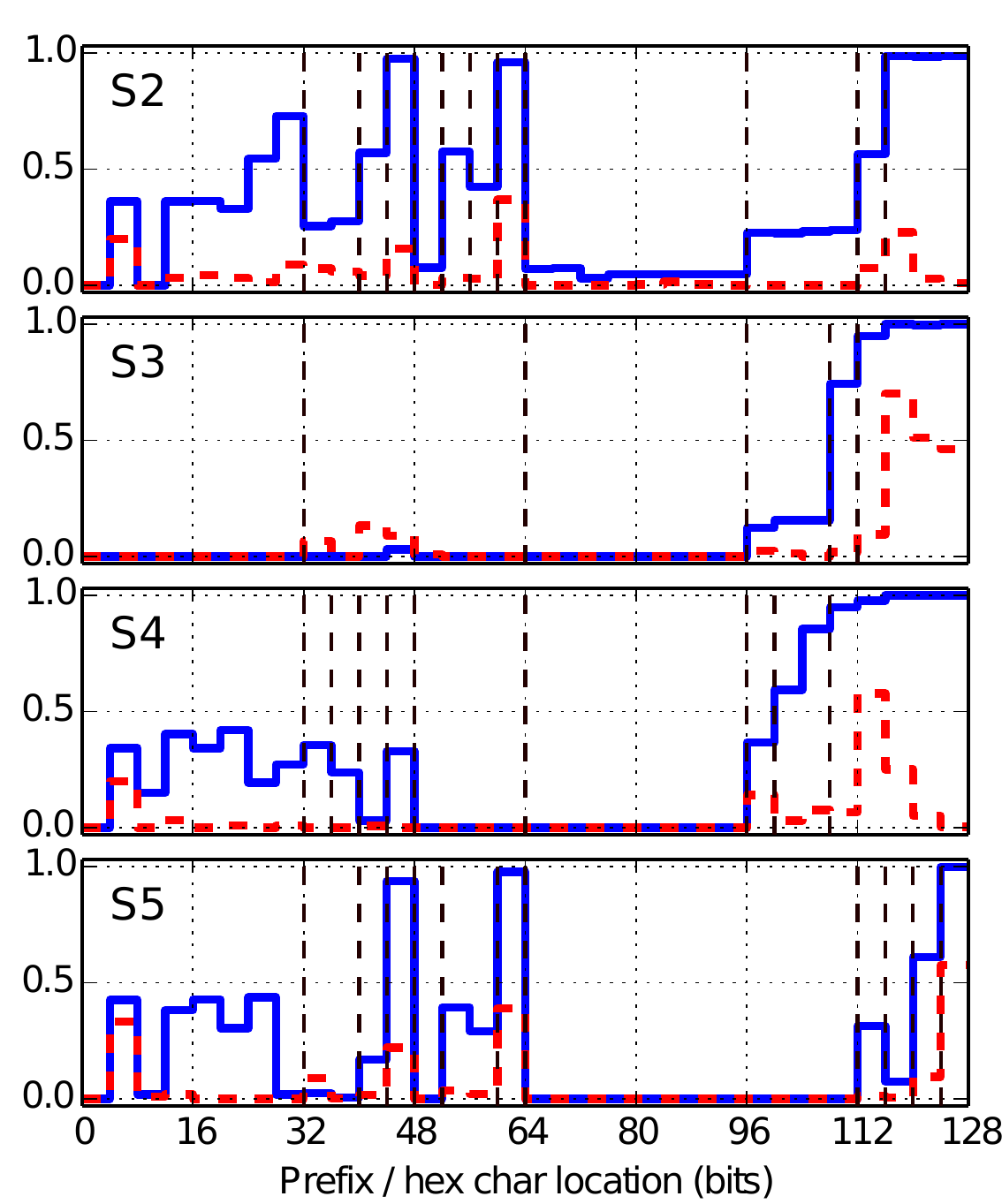} \label{fig:brief_s}} \hspace{-0.8em}
    \subfigure[routers]{\includegraphics[scale=0.51,clip=true,trim=18mm 4mm 72mm 10mm]{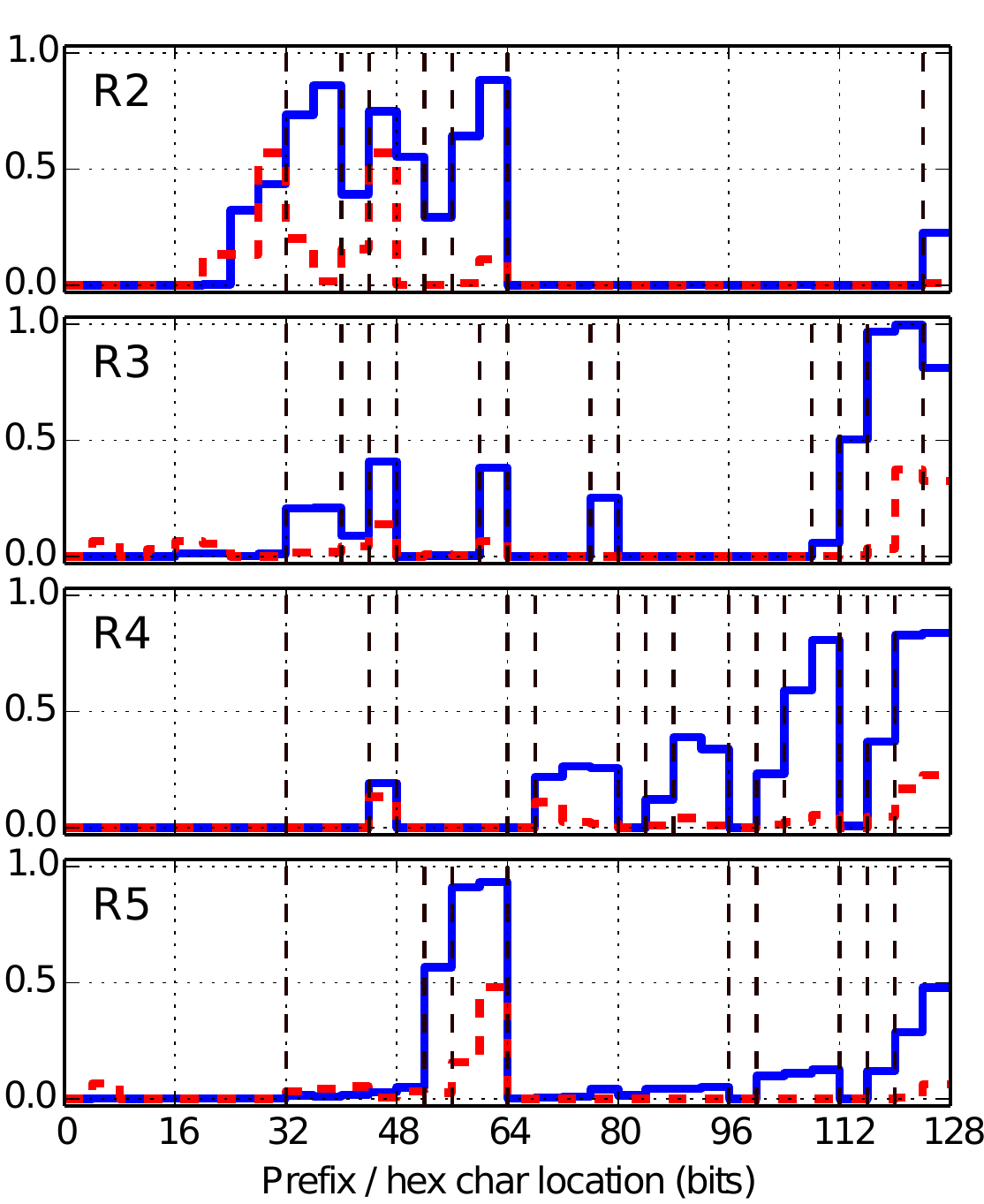} \label{fig:brief_r}} \hspace{-0.8em}
    \subfigure[clients]{\includegraphics[scale=0.51,clip=true,trim=18mm 4mm 72mm 10mm]{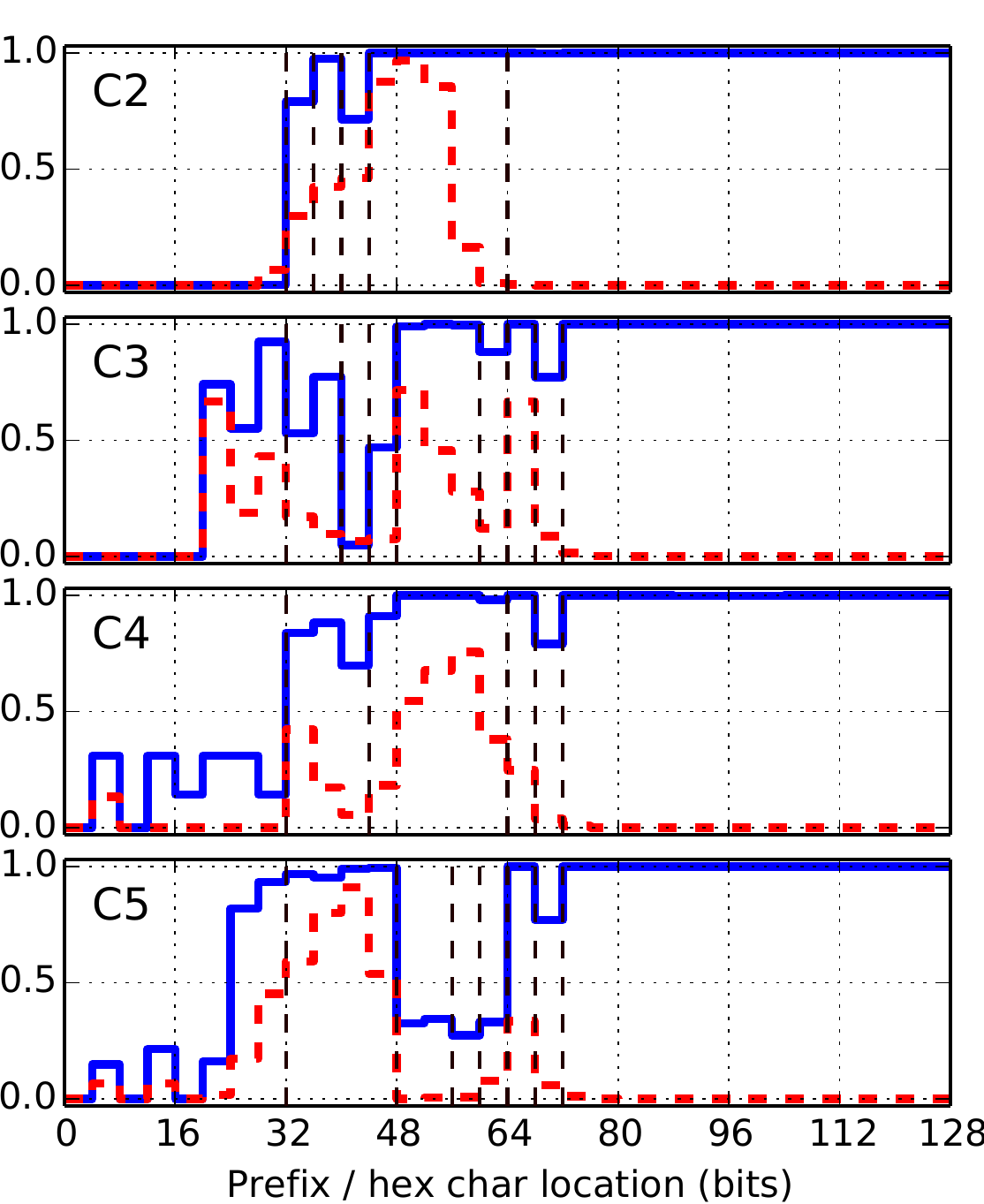} \label{fig:brief_c}}
	\caption{Brief plots for server datasets \texttt{S2-S5} (a), router datasets \texttt{R2-R5} (b), and client datasets \texttt{C2-C5} (c). Solid blue lines show per-nybble entropy and dashed red lines show 4-bit ACR. Segment labels skipped for brevity.} \label{fig:brief}
\end{figure*}
We observe a similar phenomenon for segments {\em G-J} (i.e., bottom 64 bits): each /64 prefix contains just a few IPv6 addresses. A subset of interface identifiers appears pseudo-random, \emph{e.g.}, see G14 in Table \ref{tab:analysis}. However, due to non-random addressing variants discussed above, the entropy in Fig.~\ref{fig:s1a} is below 1. Note no drop in entropy for bits 68-72 (characteristic for SLAAC), which suggests the operator has its own algorithm for generating interface identifiers. Moreover, entropy in segments {\em H-J} reveals a structure in the addresses, but without any effect on ACR, thus, unlikely to be subnetting.

Due to space constraints, we briefly present entropy analysis for the rest of server datasets in Fig.~\ref{fig:brief_s}. The addresses exhibit less variability across hex characters and ostensibly do not use SLAAC, which is consistent with our observations for {\tt AS}. Plots for \texttt{S2} and \texttt{S3} demonstrate addressing effects for different types of CDNs; the first network distributes traffic using DNS and IP unicast, while the second employs IP anycast. In consequence, \texttt{S2} has many globally distributed prefixes, while \texttt{S3} basically uses just one /96 prefix worldwide, which is reflected in entropy plots. For \texttt{S4}, a major cloud provider, we found that---apart of a simple structure in bits 32-48---only the last 32 bits are utilized for discriminating hosts and networks. For \texttt{S5}, we found that the last 2-4 nybbles often identify the service type (or type of content), deployed across many /64 prefixes (inferred by manual analysis of DNS records).
In sum, our analysis identified structure in all evaluated sets.

\subsection{Routers}
\begin{figure*}[htb]
	\centering
	\subfigure[entropy vs. 4-bit ACR]{\includegraphics[scale=0.51,clip=true,trim=10mm 68mm 16mm 5mm]{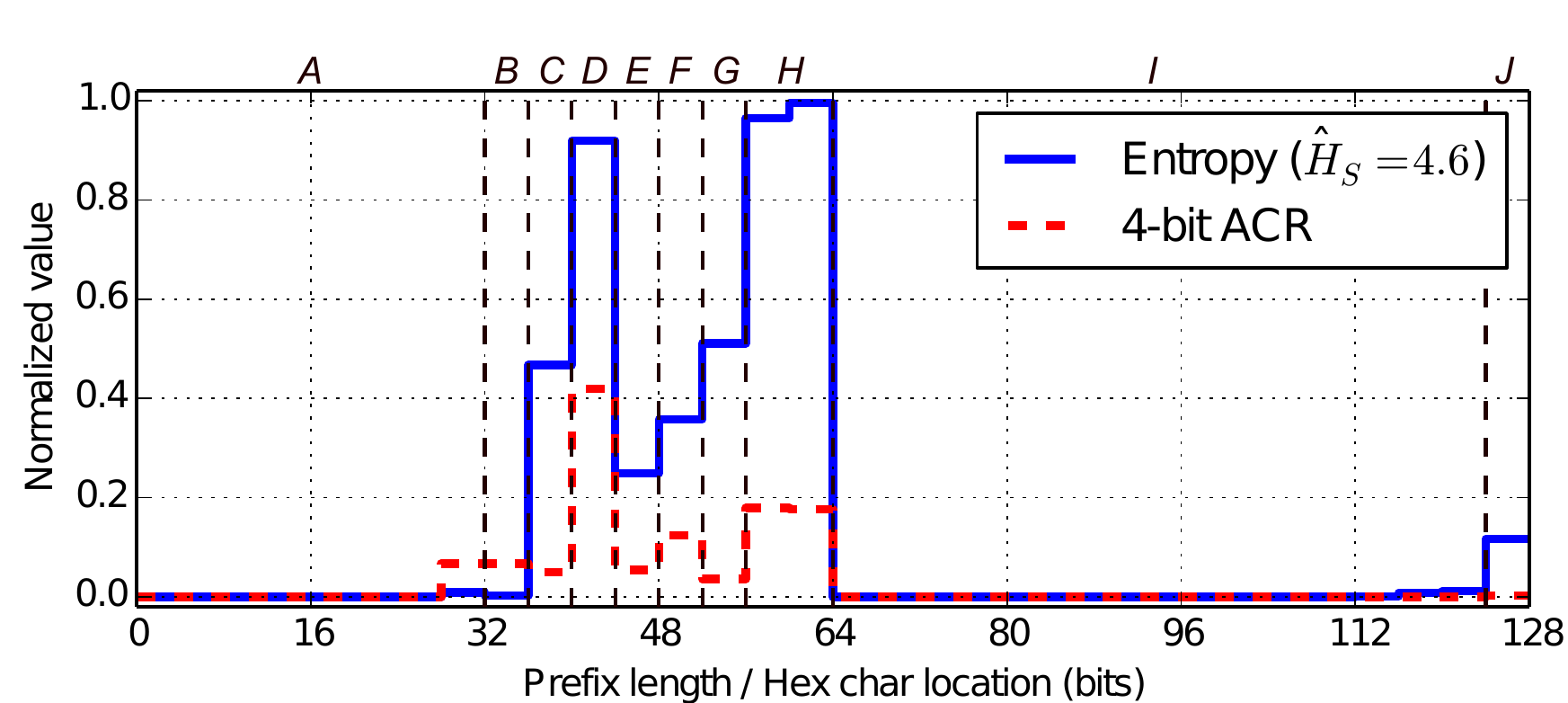} \label{fig:r1a}}
	\subfigure[BN probabilities] {\includegraphics[scale=0.88,clip=true,trim=59mm 236mm 57mm 17mm]{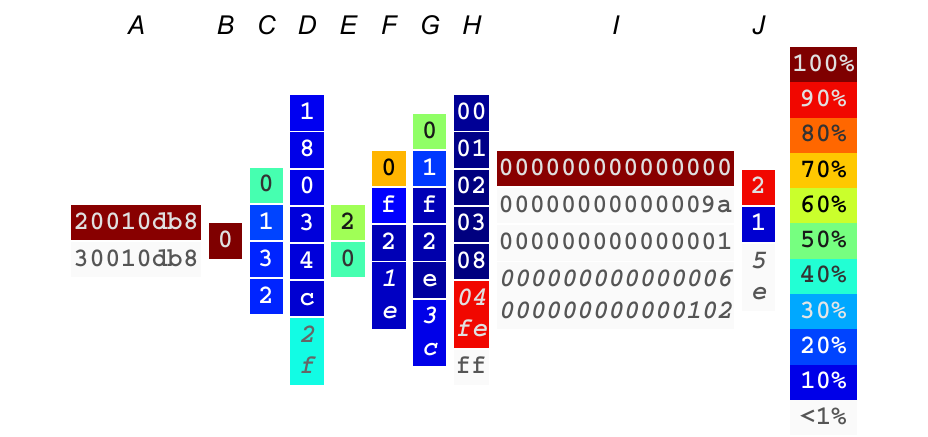} \label{fig:r1b}} \vspace{-0.5em}
     \caption{Results for router dataset \texttt{R1}.} \label{fig:r1} \vspace{-0.5em}
\end{figure*}

Fig.~\ref{fig:r1a} illustrates analysis of the \texttt{R1} dataset. We see a clear division between bits used ostensibly for discriminating prefixes (bits 28-64) and for discriminating interfaces within the prefixes (mostly bits 124-128). The network, clearly, does not implement pseudo-random interface identifiers, which is visible in entropy close to zero for bits 64-124. By evaluating the BN model, visualized in Fig.~\ref{fig:r1b}, we find that segment {\em I} is largely a string of zeros, and the last hex character is either \texttt{1} or \texttt{2}, which we believe is common for point-to-point links between this network's routers.

Consider the analysis of the other router datasets, briefly presented in Fig.~\ref{fig:brief_r}. For \texttt{R2}, we find a similar pattern as in \texttt{R1}: bottom 64 bits equal either \texttt{1} or \texttt{2}. For \texttt{R3}, we find bits 48-116 to follow a quite predictable pattern, with a majority of the hex characters equal to zero. We see that bits 32-48 discriminate prefixes, and the last 12 bits largely appear pseudo-random. However, note that values that look random do not necessarily come from a random number generator: the administrator could be systematically assigning values to these nybbles in near equal proportions. Interestingly, in \texttt{R4}, we find the IPv6 interface identifiers encode literal IPv4 addresses (ostensibly assigned to the same router interface), written as octets in base 10 across 16-bit aligned words ({\em i.e.,} colon-separated in IPv6 presentation format), hence the IID pattern visible in Fig.~\ref{fig:brief_r} for {\tt R4}. Finally, for \texttt{R5}, we find the addresses to discriminate largely in bits 52-64, while the bottom bits follow a predictable structure. Overall, we find the router addresses to implement simple patterns, yet unique and variable across operators.

\subsection{Clients} \label{sec:clients}
\begin{figure*}[htb]
    \centering
    \subfigure[entropy vs. 4-bit ACR]{\includegraphics[scale=0.51,clip=true,trim=10mm 68mm 16mm 6mm]{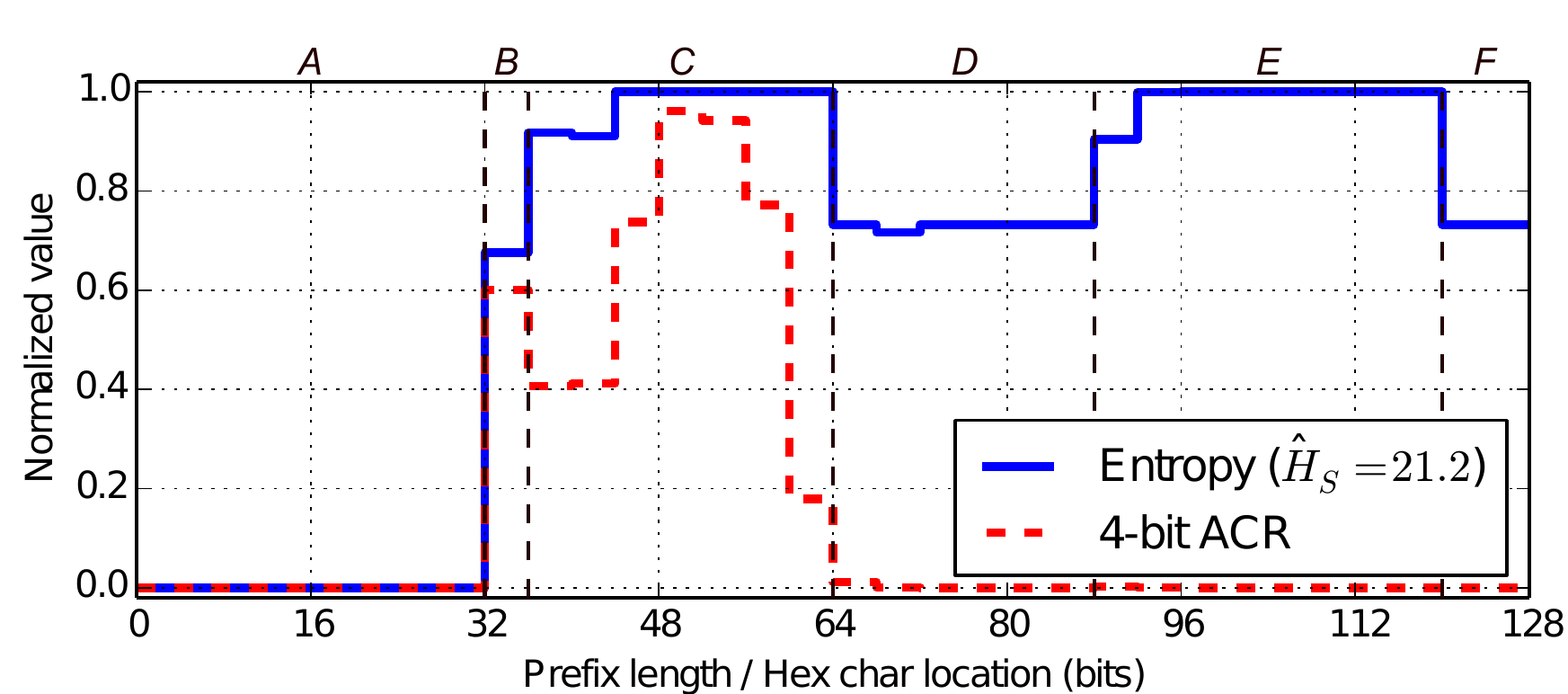} \label{fig:c1a}}
	\subfigure[BN probabilities conditional on F equal \texttt{01}]{\includegraphics[scale=0.88,clip=true,trim=60mm 238mm 58mm 17mm]{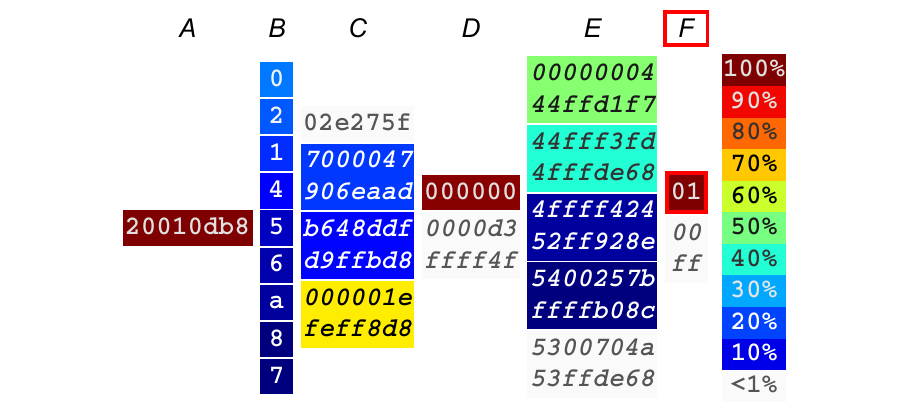} \label{fig:c1b}} \vspace{-0.5em}
     \caption{Results for client dataset \texttt{C1}.} \label{fig:c1} \vspace{-0.5em}
\end{figure*}

In Fig.~\ref{fig:c1a}, we present the analysis for dataset \texttt{C1}. The addresses correspond to a large mobile operator. Our analysis finds only six segments in the addresses, three of which uncover statistical structure in interface identifiers. In Fig.~\ref{fig:c1b}, we visualize the BN model conditioned on the last two hex characters equal to \texttt{01}, which corresponds to 47\% of all IPs in dataset \texttt{C1}.

In segments {\em B-C}, we see that bits 32-64 discriminate prefixes, as the ACR value is relatively high. Segment {\em B} takes only lower values ({\tt 0}-{\tt 8}), hence its entropy is lower than in segment {\em C}. For segments {\em D-F} we expected to see the impact of privacy addressing, already observed for the client aggregate (Fig. \ref{fig:aggregates}). Instead, we find an extraordinary pattern of entropy close to 0.7 for segments {\em D} and {\em F}, and entropy close to 1 for the segment {\em E} between them. The lower entropy in {\em D} and {\em F} is due to two popular values: \texttt{00000} ({\em D1}) and \texttt{01} ({\em F1}). Further, we find (in the BN model) that segments {\em D} and {\em F} are statistically dependent, which is visible in Fig. \ref{fig:c1b}: conditioning the BN model on {\em F1} makes {\em D} a string of zeros ({\em i.e.}, {\em D1}). We also found dependence between {\em E} and {{\em F}, which makes the first hex character in segment {\em E} more predictable when the addresses end with {\em F1}, {\em i.e.,} {\tt 00} (manifested in lower entropy for bits 88-92). On the other hand, when addresses end with other values, all bottom 64 bits are pseudo-random.

In order to investigate the underlying source of the pattern, we examined log files of the CDN corresponding to dataset \texttt{C1}. By ``User-Agent'' headers in web requests made to the CDN from that particular network, we find the addresses that follow the pattern originate only from Android devices of one popular vendor. In contrast, the addresses without the pattern originate from mobile devices of various vendors. Thus, we suspect the reason for what we see in Fig.~\ref{fig:c1b}, lies in client-side interface identifier selection in a specific version of Android built by a particular vendor for that network. We did not observe similar patterns in any other evaluated dataset.

Fig. \ref{fig:brief_c} displays entropy vs. ACR plots for the other datasets of active client addresses. For these datasets, we found the interface identifiers to be pseudo-random, {\em e.g.,} as in privacy addresses; this is visible where entropy is near 1 and ACR is near 0 in the low 64 bits. This matches our findings for dataset \texttt{AC}. However, we did not observe a drop in entropy on bits 68-72, characteristic for SLAAC privacy addresses, for dataset \texttt{C2}, which represents a mobile operator.  In all client addresses, we found various structures in bits 32-64 (or even 20-64), which demonstrates that operators implement their own addressing schemes for the /64 prefixes. Below we show some of these structures are predictable.

\subsection{Scanning IPv6 Networks}

One of the supplementary applications of our method is candidate address generation for active measurements of the IPv6 Internet. In this subsection, we evaluate Entropy/IP for scanning addresses of servers and routers.

For scanning IPv6 addresses, we apply the following methodology. A BN model for a particular network is trained on a random sample of 1K real IPv6 addresses known \emph{a priori}. Then, the model is used to generate 1M candidate targets for scanning, and the result is evaluated. For our experiments we used the datasets \texttt{S*} and \texttt{R*}: for each of them, we randomly selected 1K IPs as the training set, and used the remaining part as the testing set.

We counted the number of generated candidates that were present in the testing set, which is presented in column ``Test set'' in Table \ref{tab:scan}. Independently, for each dataset we scanned the 1M generated IPs using ICMPv6 echo requests: we show the number of IPs that replied with an echo response in column ``Ping''. Similarly, column ``rDNS'' gives the number of IPs that had a reverse DNS record. (We manually removed records that appeared dynamically generated.) In column ``Overall,'' we present the number of candidate IPs that passed at least one of the three tests, with the corresponding success rate in the next column. For these IPs, in column ``New /64s'' we give the number of /64 prefixes that were not present in the training data. Note that some networks use just a few prefixes globally (\emph{e.g.}, \texttt{S3}).

\begin{table}[tb]
\centerline{\includegraphics[scale=0.71,clip=true,trim=13mm 191mm 83mm 10mm]{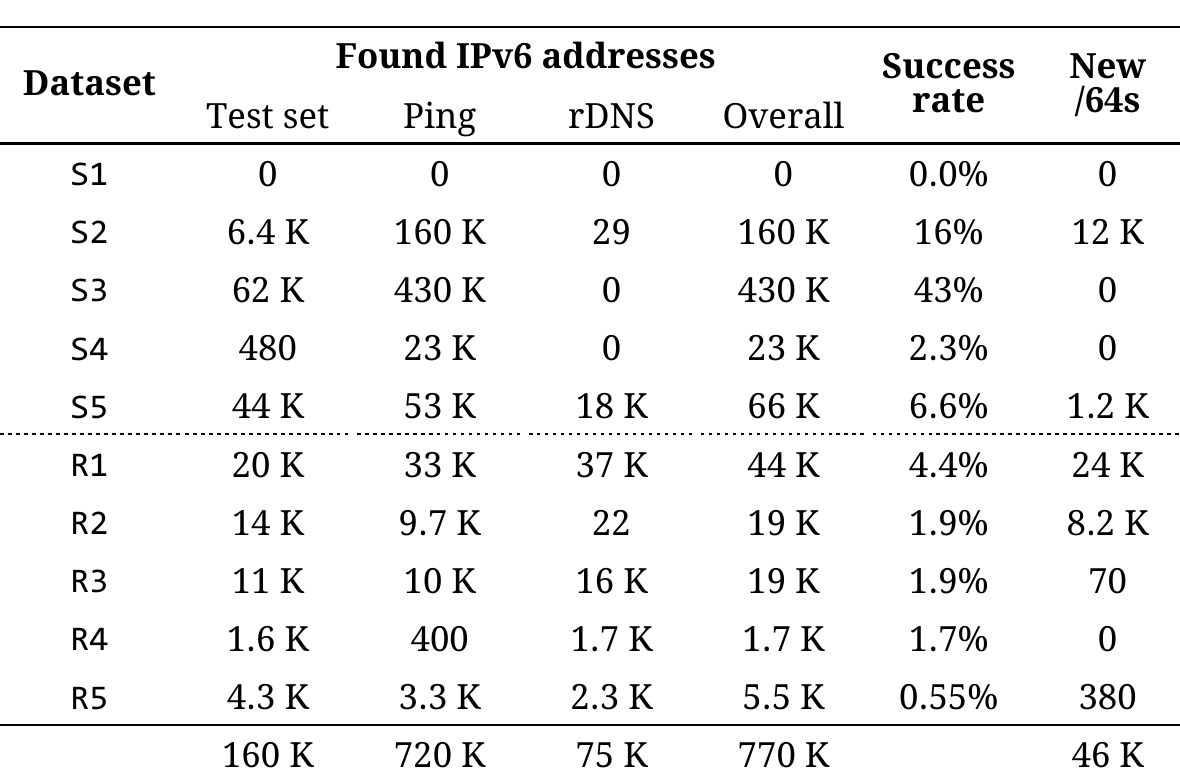}}
\vspace{-0.25em} \caption{IPv6 scanning results for Servers (\texttt{S*}) and Routers (\texttt{R*}). Models trained on 1K real addresses and tested on 1M generated addresses.} \vspace{-1em} \label{tab:scan}
\end{table}
\begin{table}[tb]
\centerline{\includegraphics[scale=0.75,clip=true,trim=10mm 237mm 105mm 5mm]{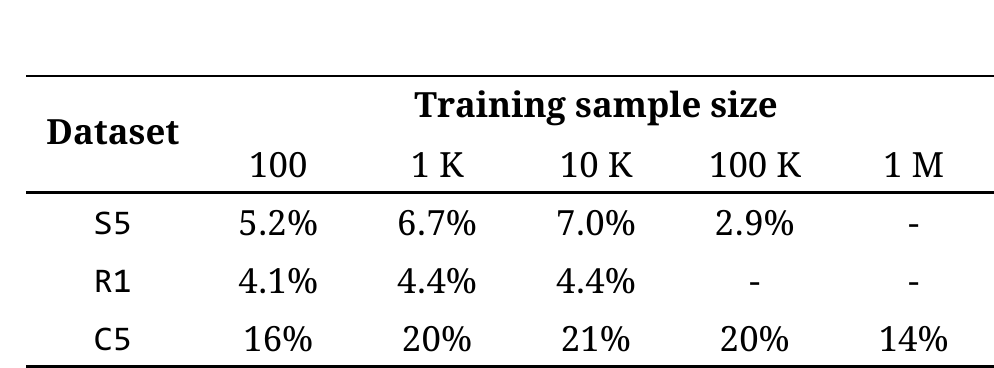}}
\vspace{-0.25em} \caption{Success rate vs. training set size.} \vspace{-1em}
\label{tab:scan_sens}
\end{table}

As visible in Table~\ref{tab:scan}, in total we found 770K addresses and 46K prefixes in 10 networks. We used only 1K training IPs per network, which corresponds to a considerable gain in discovered IPs vs.~what we used for training the BN models. We believe this resembles a relatively common situation in which one has a limited set of existing IPs from the target network and wishes to use them to bootstrap active address discovery. We found it possible to find new addresses for each dataset except for \texttt{S1}, which uses largely pseudo-random IIDs for a major subset of its addresses. On the other hand, we achieved 43\% success rate for \texttt{S3} (a large CDN). Note the probability of guessing by chance an IPv6 address in a known /32 prefix is approx. 1 in $2^{96}$.

We acknowledge some limitations of our evaluation method: first, part of the positive responses to our ``Ping'' and ``rDNS'' tests might have been generated automatically (\emph{e.g.} replying to any ping request destined to a certain prefix, causing false positives); second, the network may have far fewer hosts than the generated 1M candidates; finally, we might get a number of false negatives due to restricted datasets and due to networks blocking our ping requests and reverse DNS queries.

We believe our results indicate that Entropy/IP implements a fruitful approach to IPv6 scanning. Comparing to existing works, we find it complementary. The \texttt{scan6} tool by Gont~\cite{ipv6toolkit} and the method by Ullrich {\em et al.}~\cite{ullrich2015reconnaissance} attempt to predict interface identifiers; they do not tackle guessing the network identifiers. In contrast, our statistical model successfully predicted active /64 prefixes not seen in the training sets.

In Table \ref{tab:scan_sens} we show the effect of the training set size on the success rate for a server dataset \texttt{S5} and a router dataset \texttt{R1}. We found that a larger training set often does not cause better scanning performance and can even make it worse. We believe the reason for what we experimentally observed is that more training IPs make the BN model better at adhering to already seen data vs. generating completely new addresses.

\subsection{Predicting IPv6 Client Prefixes} \label{sec:predict_prefix}

The addresses in the sample client networks (\texttt{C1-C5}) extensively use pseudo-random interface identifiers, and thus there is no point
in trying to guess the full address. For this reason, in this subsection, we turn to predicting the prefixes, \emph{i.e.}, generating the candidates for the first 64 bits in the client addresses.

For that purpose, we constrained Entropy/IP to the top 64 bits, without any other modification to our method. We used a similar evaluation approach as in the previous subsection. We trained the model on a random 1K sample of prefixes seen on March 17th 2016 and tested them on the rest of data: for each of 1M candidate prefixes, we checked if it was observed on the same day and in the following week, as seen in the dataset.

\begin{table}[b] \vspace{-1em}
\centerline{\includegraphics[scale=0.75,clip=true,trim=10mm 221mm 115mm 10mm]{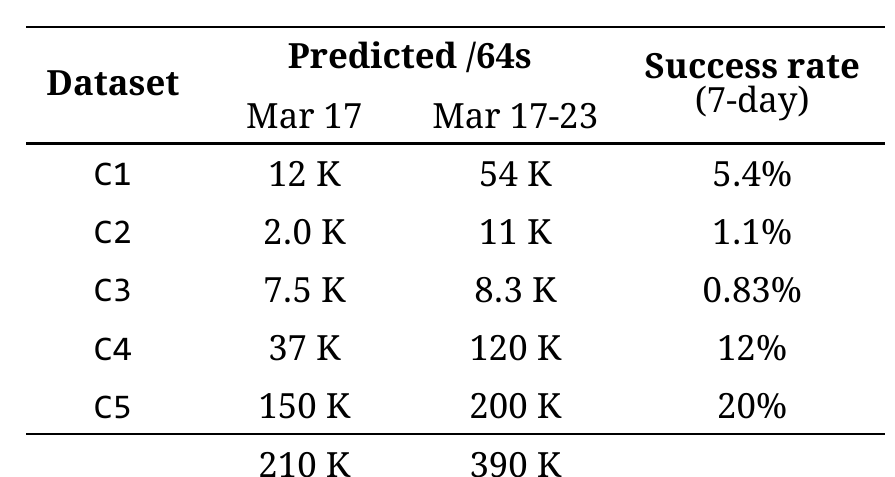}}
\caption{Prefix prediction results for Clients (\texttt{C*}), for two time spans. Models trained on 1K real /64 prefixes and tested on 1M generated prefixes.}
\label{tab:scan_cli}
\end{table}

In Table \ref{tab:scan_cli}, we present the results. We were able to predict thousands of client /64 prefixes for each network, with success rates ranging ${\sim}1\%$ to 20\%. For some datasets we found it easier to predict the prefix for the whole week vs. single day (\emph{e.g.}, \texttt{C1}), while for others we did not find such a difference (\emph{e.g.}, \texttt{C3}). We believe this demonstrates that operators have various addressing strategies for the top 64 bits of their IPv6 addresses, some of which are easy to predict. Moreover, we were able to repeat the same using the BitTorrent \texttt{AT} dataset for training. For example, by training a BN model using 1K prefixes of BitTorrent peers belonging to the network represented in \texttt{C4}, we were able to predict 120K prefixes from that dataset.

Finally, in Table \ref{tab:scan_sens} we briefly show the effect of the training sample size for \texttt{C5}. Again, using a too large training set can harm the predictive performance.

\subsection{Validation} \label{sec:validation}

To qualitatively validate our results, we prepared a briefing for each of the
five networks that we studied, at least one from each of the Servers,
Routers, and Clients categories, mentioned in Section~\ref{sec:data}.
The briefings
contained both a link to the Entropy/IP interactive user interface with an
analysis of the given network and subjective assessments based on our
interpretation.  We emailed briefings to four subject matter
experts (SME), each responsible for one of those IPv6 networks, and asked
them to confirm, refute, or otherwise comment on our results,
promising to publish them in anonymized fashion
(herein).~\footnote{\small
We were unable to identify an expert for one of the networks, even after
weeks with assistance from our network operations colleagues.}
Summarizing their feedback:

\begin{itemize} [wide]

\item The first SME, from a mobile carrier network, wrote
``In short, I don't know. {\tt :)}''
They reported that they assign prefixes to gateways from a
common vendor for service providers in the US, and these gateways in turn
assign the /64 prefixes and hint about an IID to the user
equipment, {\em i.e.,} SLAAC for 3GPP cellular hosts~\cite{3GPPSLAAC}.
Regarding a detail, they were surprised that our analysis showed
47\% of the IIDs end with {\tt 01}, {\em i.e.,} apparently
not pseudo-random. They hypothesized these addresses are
selected by some Android handsets.

\item The second SME wrote ``[Your] assumptions when it comes [to]
infrastructure are wrong.'' This SME said that internally they use a RIR block that is not routed on the Internet, and that their
routers are in a block that is only used for that purpose.
In the briefing, we guessed that the lowest IPv6 address bits sometimes
identified services. The SME confirmed this is the case, but
did not comment on many other assertions including our guesses
as to the uses of prefixes containing ``vanity'' hexadecimal
strings that spell English words in the midst of their IPv6 addresses.

\item Another network's SME wrote, ``of your analysis [...]:
nice reverse engineering of our address plan {\tt ;)}.''
They also mentioned some confusion because the initial segment {\em A} in
the conditional probability browser showed a /32 prefix that contained
some addresses that are not theirs.  This was a mistake on our part, due to
``hard-wiring'' a segment boundary at /32 and incorrectly assuming that the
entire /32 was allocated to them.  The SME asked if we used the RIR databases to verify the netblock owners.

\item The final SME responded to the details that we reported regarding
values in particular segments, saying it would require digging to
confirm or refute.
They reported that our subjective assessments
were correct, {\em e.g.,} about which prefixes were dedicated to routers
or clients, but also mentioned that was ``rather obvious.''  They
also asked questions and generously offered to discuss the results further.

\end{itemize}

With regard to the SME suggesting it may be Android handsets that
use curious IIDs in their network, independent investigation confirmed
this (Section~\ref{sec:clients}).
In answer to the question about using RIR databases, we did consult
them, but apparently not
carefully \linebreak enough.  This was useful feedback as it shows
we can improve our system by better integrating RIR data and BGP-advertised
prefixes.  With regard to some of our assessments being labeled obvious,
perhaps the SME meant they are obvious to an IPv6 operator or expert.

Overall, we find these responses interesting in their variability
with respect to how forthcoming operators are, or are not, with details
of their address plan.  We appreciate their candid, collegial responses,
such as one even admitting, essentially, that some network gear ``just
works,'' and that they did not study the resultant \linebreak SLAAC-based
address assignments to user equipment.  This comports with one of the
stated applications of our work: namely, to remotely analyze networks'
addressing practices, assess, and report potential risks.

\section{Limitations and Future Work} \label{sec:disc}

Some limitations arise from out initial assumptions and choice of
parameter values for segmentation, described in Section~\ref{sec:segments}.
First, as mentioned in Section~\ref{sec:validation}, one network's SME
identifies a problem with an
assumption we made to ``hard-wire'' a break in segments at a /32 boundary,
{\em i.e.,} after bit 32.  Sometimes this results in segments that disagree
with the actual subnet identifiers in a network's address plan.
This could, also, result from requiring segment boundaries to be between 4-bit aligned
nybbles or from algorithmic sensitivity to configurable parameters,
{\em e.g.,} the threshold and hysteresis values, which influence the
segmentation process.
While these assumptions do not necessarily prevent us from observing effects
of features smaller than 4 bits nor from observing phenomena involving
boundaries
at other bit positions, they do affect the resulting structure and candidate
subnets. In future work, we might consider removing some of these fixed
parameters or searching the parameter space with the hope of producing
structures that match real subnet boundaries, {\em e.g.,} ground truth
from operators.

Note that our Bayesian Network model captures dependencies \emph{between} segments, and that
we did not study dependencies across nybbles \emph{within} segments. We intend to do so in future research,
possibly employing the concept of \emph{mutual information}, or an entropy measure of the string of nybbles within a segment, where the normalization considers the length of that segment.

In this work, we have ignored the temporal characteristics of address
sets, treating them as if they are a set of active addresses at
one point in time.
However, future work would benefit from integrating
temporal considerations into our method, for instance with the hope
of uncovering boundaries of sequential and random assignments of
addresses from dynamic pools that we discovered in
some networks~\cite{plonka2015temporal}.
Another consideration for future work is structural analysis in time-series, {\em e.g.,} to detect changes in network deployments.

Lastly, our evaluation supports our claim that
Entropy/IP is effective at generating hit lists of candidate target
addresses if one wishes to conduct a survey or census of the IPv6 Internet by active scans. This is an obvious choice for future work that we plan to pursue.
Here, note that our tool can be used on datasets much smaller than the large sets we described in Section \ref{sec:data} and employed for evaluation. Thus, datasets known to the research community can be used for further discovery of IPv6 server farms, routing infrastructure, as well as client addresses of selected networks.

\section{Conclusion} \label{sec:concl}
Comprehensive understanding of address structure is more difficult with IPv6
than IPv4. This is due to the features and the freedom that IPv6 offers in
address assignment.  To accelerate our IPv6 investigations,
we developed a system---Entropy/IP---that automates
network structure discovery
by deeply analyzing sets of sample addresses gleaned
by standard means, {\em e.g.,} server logs, passive DNS, and {\tt traceroute}.
We demonstrate the system's effectiveness in discovering structure in both
interface identifiers {\em and} network identifiers, {\em i.e.,} subnets.
While there is future work still to do, our initial performance
evaluation suggests there is potential to surgically survey the vast Internet address space,
as well as improve analysts' understanding in operation and defense of our increasingly
complicated Internet.

\section*{Acknowledgements}
We thank Keung-Chi ``KC'' Ng, Eric Vyncke, and the four network operators who replied to our solicitation for comments.
We thank Johanna Ullrich for helpful discussion on their methods and datasets. We also express gratitude to our proof-readers:
Jan Galkowski, Steve Hoey, Grzegorz Karch, Mariusz S\l{}abicki, and our shepherd, Walter Willinger.

\newpage

\bibliographystyle{abbrv}
\bibliography{bibliography}

\end{document}